\RequirePackage{lineno}
\documentclass[prd,twocolumn,showpacs,tightenlines,superscriptaddress,showpacs]{revtex4-1}
\usepackage{amssymb}
\usepackage{amsmath}
\usepackage{graphicx}
\usepackage{subfigure}
\usepackage{epsfig}
\usepackage{dcolumn}
\usepackage{bm,color}
\usepackage{multirow}
\usepackage{verbatim}
\usepackage{float}
\usepackage{overpic}
\usepackage{soul}

\newcommand{\piz}{\pi^0}
\newcommand{\kp}{K^+}
\newcommand{\km}{K^-}
\newcommand{\ks}{K_S^0}
\newcommand{\kstar}{K^*(892)}
\newcommand{\kstarc}{K^{*}(892)^{\pm}}
\newcommand{\kstarn}{K^{*}(892)^{0}}
\newcommand{\etap}{\eta^{\prime}}
\newcommand{\psp}{\psi(3686)}
\newcommand{\jpsi}{J/\psi}
\newcommand{\ccbar}{c\bar{c}}
\newcommand{\ssbar}{s\bar{s}}
\newcommand{\EE}{e^+e^-}
\newcommand{\pp}{\pi^+\pi^-}
\newcommand{\kk}{K^+K^-}
\newcommand{\ksp}{K_S^0\pi^\pm}
\newcommand{\kpi}{K^\pm\pi^\mp}
\newcommand{\GG}{\gamma\gamma}
\newcommand{\ra}{\rightarrow}
\newcommand{\kskp}{K^{0}_S K^{\pm}\pi^{\mp}}
\newcommand{\kkp}{K^{+} K^{-}\pi^{0}}
\newcommand{\kkbarp}{K\bar{K}\pi}
\newcommand{\kstark}{K^*(892)\bar{K}}
\newcommand{\kstarkc}{K^{*}(892)^{\pm}K^{\mp}}
\newcommand{\kstarkn}{K^{*}(892)^{0}\bar{K^0}}
\newcommand{\ccj}{\chi_{cJ}}
\newcommand{\ccz}{\chi_{c0}}
\newcommand{\cco}{\chi_{c1}}
\newcommand{\cct}{\chi_{c2}}
\newcommand{\beq}{\begin{equation}}
\newcommand{\eeq}{\end{equation}}
\newcommand{\bitm}{\begin{itemize}}
\newcommand{\eitm}{\end{itemize}}
\newcommand{\mev}{\mathrm{MeV}}

\newcommand{\mevcc}{\mathrm{MeV}/c^2}
\newcommand{\gev}{\mathrm{GeV}}
\newcommand{\gevc}{\mathrm{GeV}/c}
\newcommand{\gevcc}{\mathrm{GeV}/c^2}
\newcommand{\sig}{\mathrm{sig}}
\newcommand{\bkg}{\mathrm{bkg}}

\begin{document}

\title{\boldmath Study of $\chi_{cJ}$ decaying into $\phi$ $\kstar$ $\bar{K}$}

\author{
  \begin{small}
    \begin{center}
      M.~Ablikim$^{1}$, M.~N.~Achasov$^{9,a}$, X.~C.~Ai$^{1}$,
      O.~Albayrak$^{5}$, M.~Albrecht$^{4}$, D.~J.~Ambrose$^{44}$,
      A.~Amoroso$^{48A,48C}$, F.~F.~An$^{1}$, Q.~An$^{45}$,
      J.~Z.~Bai$^{1}$
% ...
      \\
      \vspace{0.2cm}
      (BESIII Collaboration)\\
      \vspace{0.2cm} {\it
        $^{1}$ Institute of High Energy Physics, Beijing 100049, People's Republic of China\\
% ...
$^{4}$ Bochum Ruhr-University, D-44780 Bochum, Germany\\
        $^{5}$ Carnegie Mellon University, Pittsburgh, Pennsylvania 15213, USA\\
% ...
$^{9}$ G.I. Budker Institute of Nuclear Physics SB RAS (BINP), Novosibirsk 630090, Russia\\
% ...
$^{44}$ University of Rochester, Rochester, New York 14627, USA\\
        $^{45}$ University of Science and Technology of China, Hefei 230026, People's Republic of China\\
% ...
$^{48}$ (A)University of Turin, I-10125, Turin, Italy; (B)University of Eastern Piedmont, I-15121, Alessandria, Italy; (C)INFN, I-10125, Turin, Italy\\
% ...
$^{a}$ Also at the Novosibirsk State University, Novosibirsk, 630090, Russia\\
% ...
      }\end{center}
    \vspace{0.4cm}
  \end{small}
}




\begin{abstract}
Using a data sample of 106 million $\psp$ events collected with the BESIII
detector operated at the BEPCII storage ring, we study for the first
time the decay $\ccj\to\phi\kskp$
and $\ccj\to\phi\kkp$ in the E1 radiative transition $\psp\to\gamma\ccj$.
The decays are dominated by the three-body decay $\ccj\to \phi \kstark$.
We measure branching fractions for this reaction via the neutral and charged $\kstar$
and find them consistent with each other within the expectation of isospin symmetry.
In the $\kkbarp$ invariant mass distribution a structure near the
$\kstark$ mass threshold is observed, and the corresponding
mass and width are measured to be $1412\pm4(\mathrm{stat.})\pm8(\mathrm{sys.})~\mevcc$
and $\Gamma$ = $84\pm12(\mathrm{stat.})\pm40(\mathrm{sys.})~\mev$, respectively. The observed
state favors an assignment to the $h_1(1380)$, considering its possible $J^{PC}$ and comparing
its mass, width and decay mode to those reported in the Particle Data
Group.
\end{abstract}


\maketitle

\section{\label{sec:intro}Introduction}

It is well known that the heavy-quark mass provides a natural boundary between the perturbative and
non-perturbative regimes. Quarkonium systems are regarded as a unique laboratory to study the interplay
between perturbative and nonperturbative effects in Quantum
Chromodynamics (QCD). Exclusive quarkonium decays constitute an
important basis for investigating these effects.
Unlike the $\jpsi$ and $\psp$, the P-wave charmonia states $\ccj$ ($J$ = 0, 1, 2) are not
directly produced in $\EE$ collisions, thus are less well understood
to date~\cite{pdg}.
Obtaining more experimental data on exclusive decays of these $\ccj$ states is important for
a better understanding of their nature and decay mechanisms, as well as for testing QCD
based calculations.
Exclusive charmonium decays have been investigated widely within QCD. The
dominant dynamical mechanism is $\ccbar$ quark annihilation into the minimal number
of gluons allowed by symmetries followed by the creation of light quark-antiquark
pairs, which form the final state hadrons~\cite{hqphys}.
The $\ccj$ states are expected to annihilate into two gluons. Predictions by the color singlet
model give smaller decay widths than those determined
experimentally~\cite{csm1, csm2, csm3}, while much better predictions
can be obtained if the color octet state is taken into account~\cite{com1, com2}.
Since the $\ccj$ states are produced copiously in the E1 radiative transition of
$\psp$, with branching fractions around 9\%~\cite{pdg}, the large ${\psp}$ data sample
taken with the Beijing Spectrometer (BESIII) located at the Beijing Electron-Positron
Collider (BEPCII) provides a unique opportunity for detailed studies of $\ccj$
exclusive decays.

In the quark model, 22 $\ssbar$ sector resonances, collectively called strangeonia, are expected
below 2.2 $\gevcc$. So far only 7 states are widely accepted experimentally,
counting the maximally mixed $\eta-\etap$ as one $\ssbar$ state~\cite{ssbar}.
The axial-vector candidate, $h_1(1380)$, is assigned as the $s\bar{s}$ partner of the
1$^1P_1$ states, $h_1(1170)$, considering its mass and dominant decay to the
$K^*(892)K$ final state.
Experimentally, $h_1(1380)$ has been reported in a Partial Wave Analysis (PWA) only by LASS~\cite{lass} and Crystal
Barrel~\cite{crystalball}. The nature of this state is still controversial with respect to the predictions made
by considering the mixing between $SU(3)$-singlet and $SU(3)$-Octet mesons in the $1^{3}P_1$ and $1^{1}P_1$ nonets~\cite{DMLi}
or those made by a relativized quark model~\cite{Godfrey}. The mass
determined by the LASS measurement is significantly smaller than the theory
prediction. If the LASS result is confirmed, the $h_1(1380)$ would
seem too light to be the $1^{1}P_1$ $\ssbar$  member. The Crystal
Barrel results are consistent with theory predictions, which means
that $h_1(1380)$ could be a convincing candidate to be the $\ssbar$ partner of the $1^{1}P_1$ state $h_1(1170)$.
The measurement of the total width of the $h_1(1380)$ is thought to be complicated by the nearby $\kstark$ threshold,
where the mass distribution and effective width can not be well described with a traditional Breit-Wigner form.
The direct observation of the $h_1(1380)$ in experiments and the precise measurement of its
resonance parameters may shed light on its nature and aid in
identifying the ground state axial-vector meson nonet in the quark model.
Due to conservation of angular momentum and parity, the axial-vector strangeonia candidates are not
produced in $\jpsi$ ($\psp$) radiative decays, but are expected to be produced through the
hadronic decay of $\ccj$ associated with a vector meson $\phi$ or in
$\jpsi$ ($\psp$) decays with a pseudoscalar meson $\etap$ ($\eta$).
In this paper, we report the first measurement of the decay $\ccj\to\phi\kskp$ and
$\ccj\to\phi \kkp$ in the electric dipole (E1) radiative transition $\psp\to\gamma\ccj$.
In the following text, if not specified, $\kstark$ denotes $\kstarkn$
and its isospin-conjugate state $\kstarkc$, while $\kkbarp$ denotes both $\kskp$ and $\kkp$.
The charge conjugated channel is always implied.
This analysis is based on a data sample of $1.06 \times 10^{8}$ $\psp$ events collected
with the BESIII detector at the BEPCII.
Data with an additional integrated luminosity of 44.5
$\mathrm{pb}^{-1}$~\cite{conti} at a center-of-mass energy
of $\sqrt{s} = 3.65 ~\gev$ are used to study continuum contributions.

\section{\label{sec:detector}BESIII Detector}
\label{secii}
The BESIII detector, described in detail in Ref.~\cite{bes3}, has an effective geometrical
acceptance of 93\% of 4$\pi$.
It contains a small cell helium-based main drift chamber (MDC) which provides momentum
measurements of charged particles;
a time-of-flight system (TOF) based on plastic scintillator which helps to identify
charged particles;
an electromagnetic calorimeter (EMC) made of CsI(Tl) crystals used to measure
the energies of photons and provide trigger signals;
and a muon system (MUC) made of Resistive Plate Chambers (RPC) located
in the iron flux return yoke of the superconducting solenoid.
The momentum resolution of the charged particles is $0.5$\% at $1~\gevc$ in a 1~Tesla
magnetic field.
The energy loss ($dE/dx$) measurement provided by the MDC has a resolution better than 6\% for
electrons from Bhabha scattering.
The time resolution of the TOF is $80$~ps ($110$~ps) in the barrel (endcaps).
The photon energy resolution is $2.5$\% ($5$\%) at $1~\gev$ in the barrel
(endcaps) of the EMC.

Monte Carlo~(MC) simulated events are used to determine the detection efficiency, optimize
the selection criteria, and study possible backgrounds.
A GEANT4-based~\cite{geant4} MC simulation software, which includes the geometric and material
descriptions of the BESIII detector, the detector response, and digitization models as well as
the detector running conditions and performance, is used to generate MC samples.
The $\psp$ resonance is simulated with the \textsc{kkmc}~\cite{KKMC} generator, which is an
event generator based on precise predictions of the Electroweak Standard Model for the process
$e^{+}  e^{-} \ra f\overline{f} + n\gamma, f = \mu, \tau, d, u, s, c, b$.
The beam energy spread and initial state radiation (ISR) are taken
into account in the simulation.
The subsequent decay processes are generated with {\sc EvtGen}~\cite{EvtGen}.
Background studies are based on a sample of $10^8$ $\psp$ inclusive decays, generated with the
known branching fractions taken from the the Particle Data Group
(PDG)~\cite{pdg}, or with {\sc lundcharm}~\cite{Lundcharm} for the
unknown decays.

\section{Event selection}

Charged particles are reconstructed from hits in the MDC.
Charged tracks are required to be within the acceptance of the MDC, satisfying $|\cos\theta|<0.93$.
For each track, the point of closest approach to the interaction point (IP) must be within 1~cm
in the plane perpendicular to the beam direction and within $\pm$10~cm along the beam direction.
Particle identification (PID) is carried out by combining information from the MDC and TOF.
PID probabilities ($prob(i)$) are calculated for each track according to different particle hypotheses
$i$ ($i$ = $\pi$, $K$ and $p$). To be identified as a kaon, a track is required to have $prob(K)>prob(\pi)$
and $prob(K)>prob(p)$, while pion candidates are required to satisfy $prob(\pi)>prob(K)$ and
$prob(\pi)>prob(p)$.

Photon candidates are reconstructed from isolated showers in the EMC. Each photon candidate is required to
have a minimum energy of 25~$\mev$ in the EMC barrel region ($|\cos\theta|<0.8$) or 50~$\mev$
in the endcap region ($0.86<|\cos\theta|<0.92$).
In order to improve the reconstruction efficiency and the energy resolution, the energy deposited
in the nearby TOF counters is included in the photon reconstruction.
The timing information from the EMC is used to further suppress electronic noise and energy deposition
unrelated to the event of interest.

$\ks$ candidates are reconstructed with all combinations of two oppositely charged tracks
(without a requirement on the point of closest approach to the IP), assuming both tracks to be pions.
A secondary vertex fit is performed for each combination.
Each $\ks$ candidate is required to have an invariant mass that satisfies $|M_{\pp}-M_{\ks}|<10~\mevcc$
and a decay length two times larger than its fit error, where $M_{\ks}$ is
the nominal mass of $\ks$ taken from the PDG~\cite{pdg}.
If more than one $\ks$ is reconstructed within an event, the one with the minimum
$|M_{\pp}-M_{\ks}|$ is selected for further analysis.

$\piz$ candidates are reconstructed from pairs of photons whose invariant mass satisfies
$[M_{\piz}-60] <M_{\GG} < [M_{\piz}+40]~\mevcc$, where $M_{\piz}$ is the nominal
mass of $\piz$ taken from the PDG~\cite{pdg}. An asymmetrical mass window is used for $\piz$ reconstruction because
the photon energy deposited in the EMC has a long tail on the low energy side.
A kinematic fit is performed on the selected
photon pairs by constraining their invariant mass to the $\piz$ mass
($1C$ fit). The $\chi^2_{1C}$
from this kinematic fit is required to be less than 25.
To remove backgrounds in which the $\piz$ is falsely reconstructed from a high energy 
photon paired with a spurious shower, a restriction is imposed on the decay angle $|\cos\theta_\mathrm{decay}|<0.95$,
where $\theta_\mathrm{decay}$ is the polar angle of each photon in the $\piz$ rest frame
with respect to the $\piz$ direction in the $\psp$ rest frame.
If more than one $\piz$ is found within an event, the one with the minimum $|M_{\GG}-M_{\piz}|$
is selected for further analysis.

In the selection of the decay chain $\psp \to \gamma \ccj$, $\ccj\to\phi \kskp$, $\phi \to \kk$, a candidate
event is required to contain a $\ks$ candidate, exactly four additional charged
tracks with zero net charge, and at least one photon.
The four additional charged tracks must be identified as three kaons and one pion according to PID information.
In the selection of $\psp \to \gamma \ccj$, $\ccj\to\phi\kkp$, $\phi\to\kk$, a candidate event
is required to have four charged tracks with zero net charge, one $\piz$ candidate, and at least one additional photon.
The four charged tracks must be identified as two positively charged and two negatively
charged kaons, respectively.

To further remove potential backgrounds and to improve the mass resolution, a four-constraint
energy-momentum conservation kinematic fit ($4C$ fit) is performed. Events in the reaction $\psp\to\gamma
\kk \kskp$ ($\psp\to\gamma \kk\kkp$) are required to have a $\chi^2_{4C}
< 100$ ($\chi^2_{4C} < 40$).
For events with more than one photon (besides the photons from the $\piz$ decay in the $\ccj\to \phi
\kkp$ channel), the $4C$-fit is repeated with each photon candidate. The photon candidate which gives the minimum
$\chi^2$ is selected to be the radiative photon from the $\psp$ decay.

In the selection of $\psp\to\gamma\kk\kkp$ events, an additional requirement, $|M_{\kk\kk}-M_{\jpsi}|>30
~\mevcc$, is imposed to suppress the backgrounds $\psp \to \piz\piz\jpsi$ and $\jpsi \ra \kk\kk$,
where $M_{\jpsi}$ is the nominal mass of $\jpsi$ taken from the PDG~\cite{pdg}.

\begin{figure}[htbp]
    \begin{center}
    \begin{overpic}[width=0.45\textwidth,angle=0]{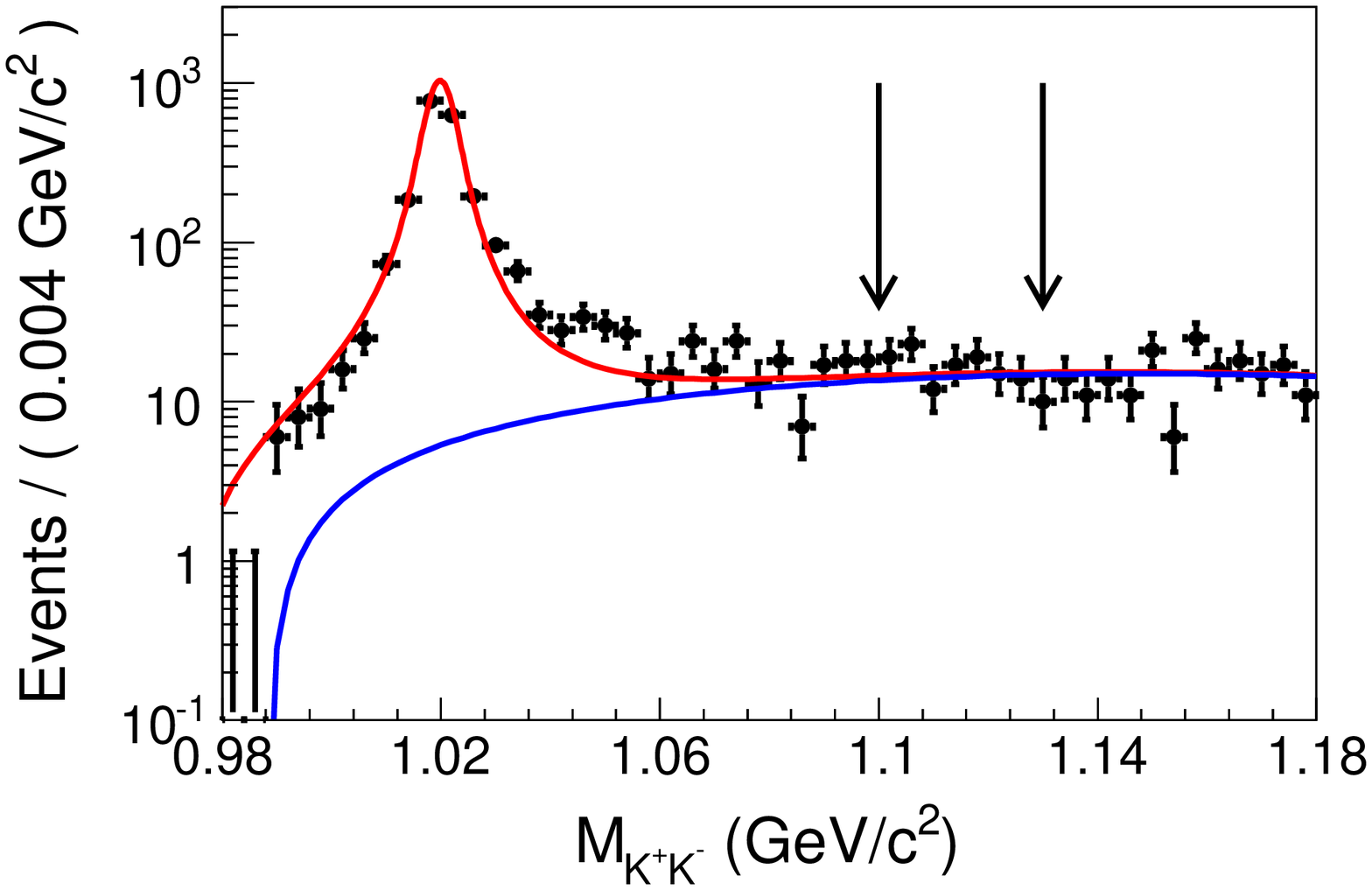}
    \put(45,50){(a)}
    \end{overpic}
    \begin{overpic}[width=0.45\textwidth,angle=0]{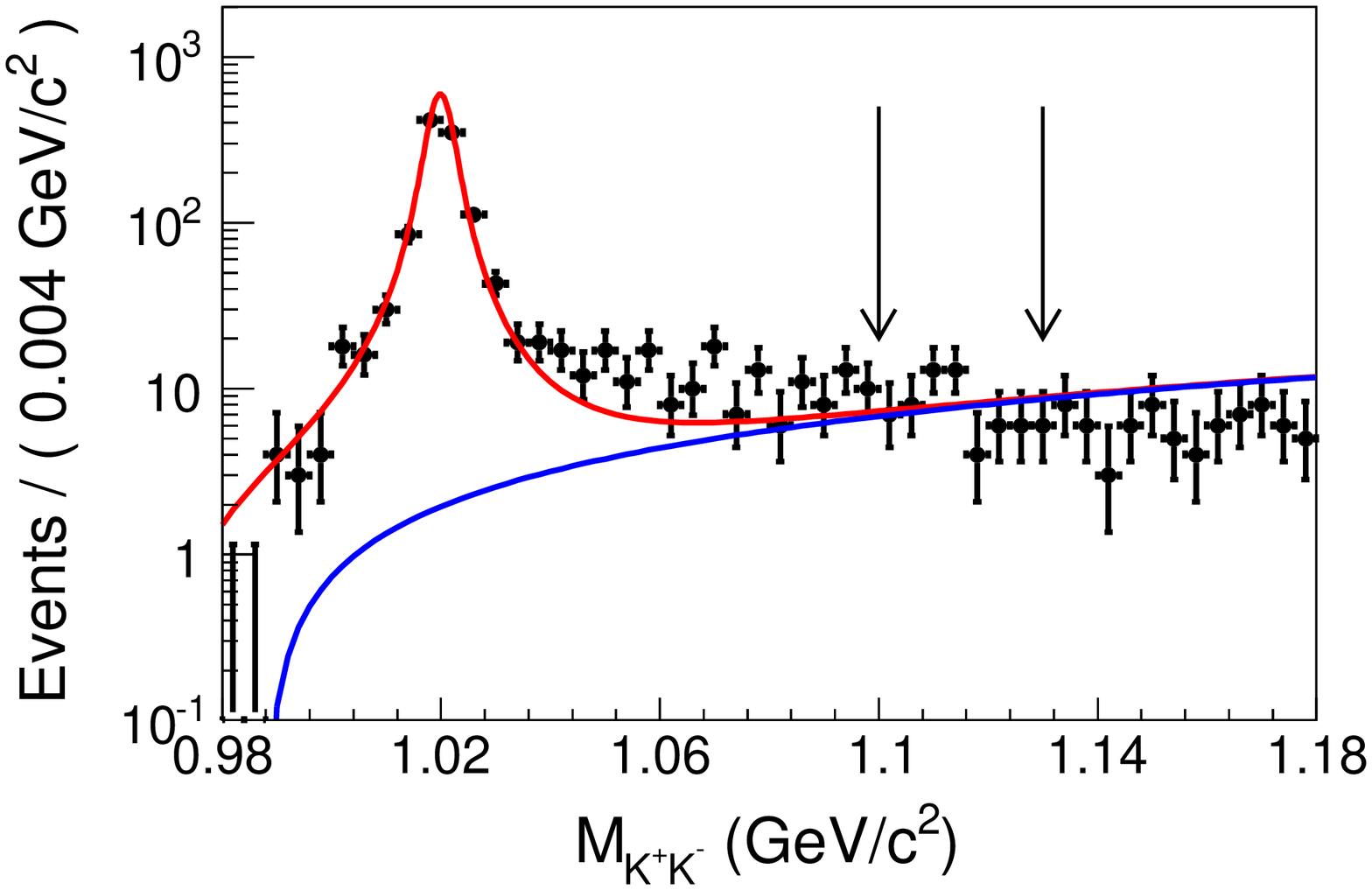}
    \put(45,50){(b)}
    \end{overpic}
    \end{center}
    \caption{%
        Invariant mass distribution of $\kk$ for $\phi$ candidates
        (a) for the candidate events of $\gamma \kk\kskp$,
        (b) for the candidate events of $\gamma \kk\kkp$.
        The arrows indicate the $\phi$ sideband region.}%
    \label{fig:phimass}
\end{figure}
After the above selection criteria are applied, the
decay $\phi \to \kk$ is reconstructed using the
two oppositely charged kaons whose invariant mass is closest to the nominal $\phi$ mass.
Fig.~\ref{fig:phimass} shows the $\kk$ invariant mass of $\phi$ candidates for events in which
the invariant mass of $\kk\kskp$ ($\kk\kkp$) is between 3.35 and 3.6 $\gevcc$. The $\phi$ signal is described by P-wave relativistic Breit-Wigner function, and the background is a 3rd order Chebyshev polynominal function.
A clear $\phi$ signal is observed above a very low background. Signal events are extracted by applying
a mass window requirement, $|M_{\kk}-M_{\phi}|<15~\mevcc$, which corresponds to three times the mass resolution.

\begin{figure}[htbp]
    \begin{center}
    \begin{overpic}[width=0.45\textwidth,angle=0]{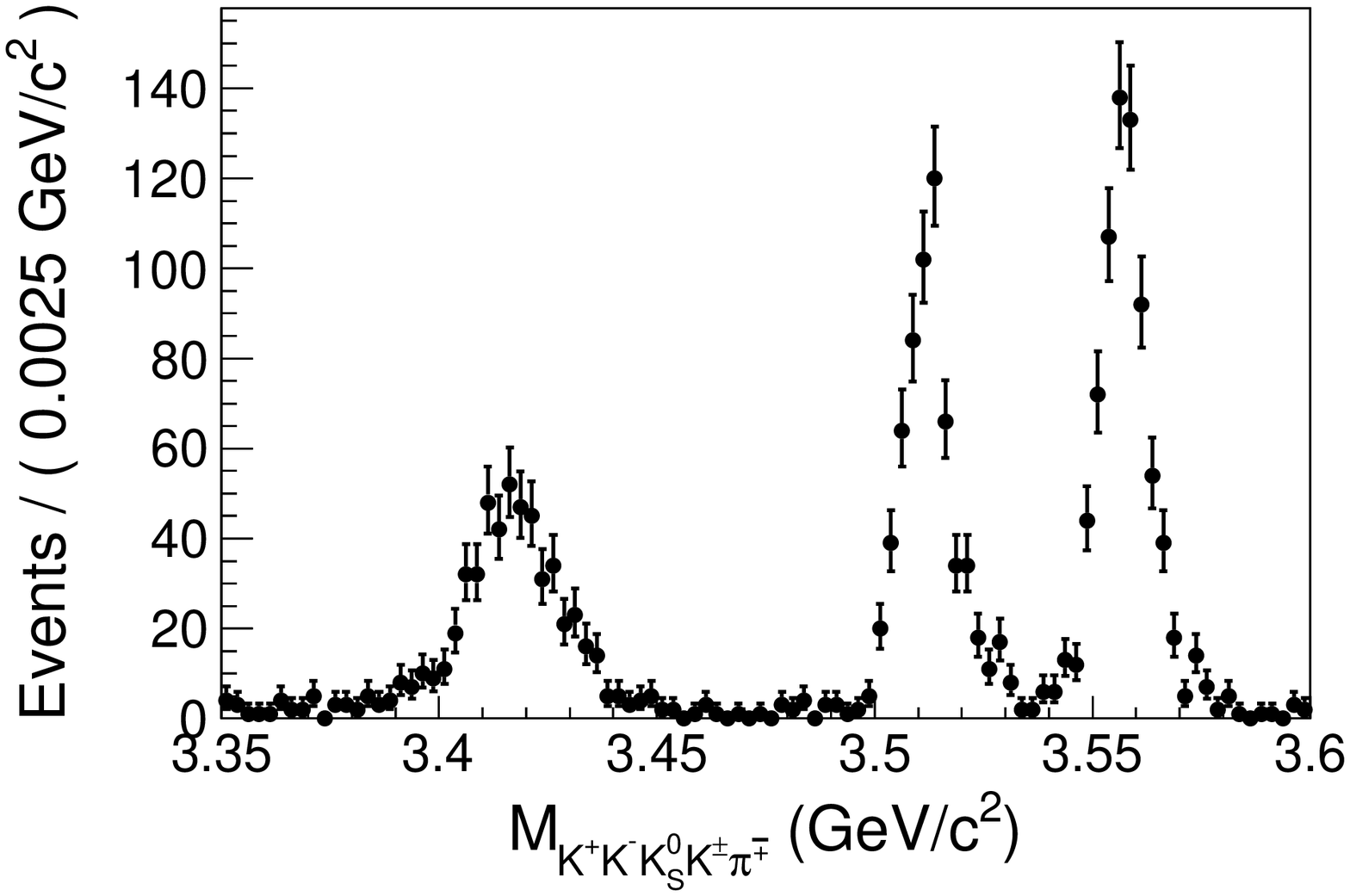}
    \put(25,50){(a)}
    \end{overpic}
    \begin{overpic}[width=0.45\textwidth,angle=0]{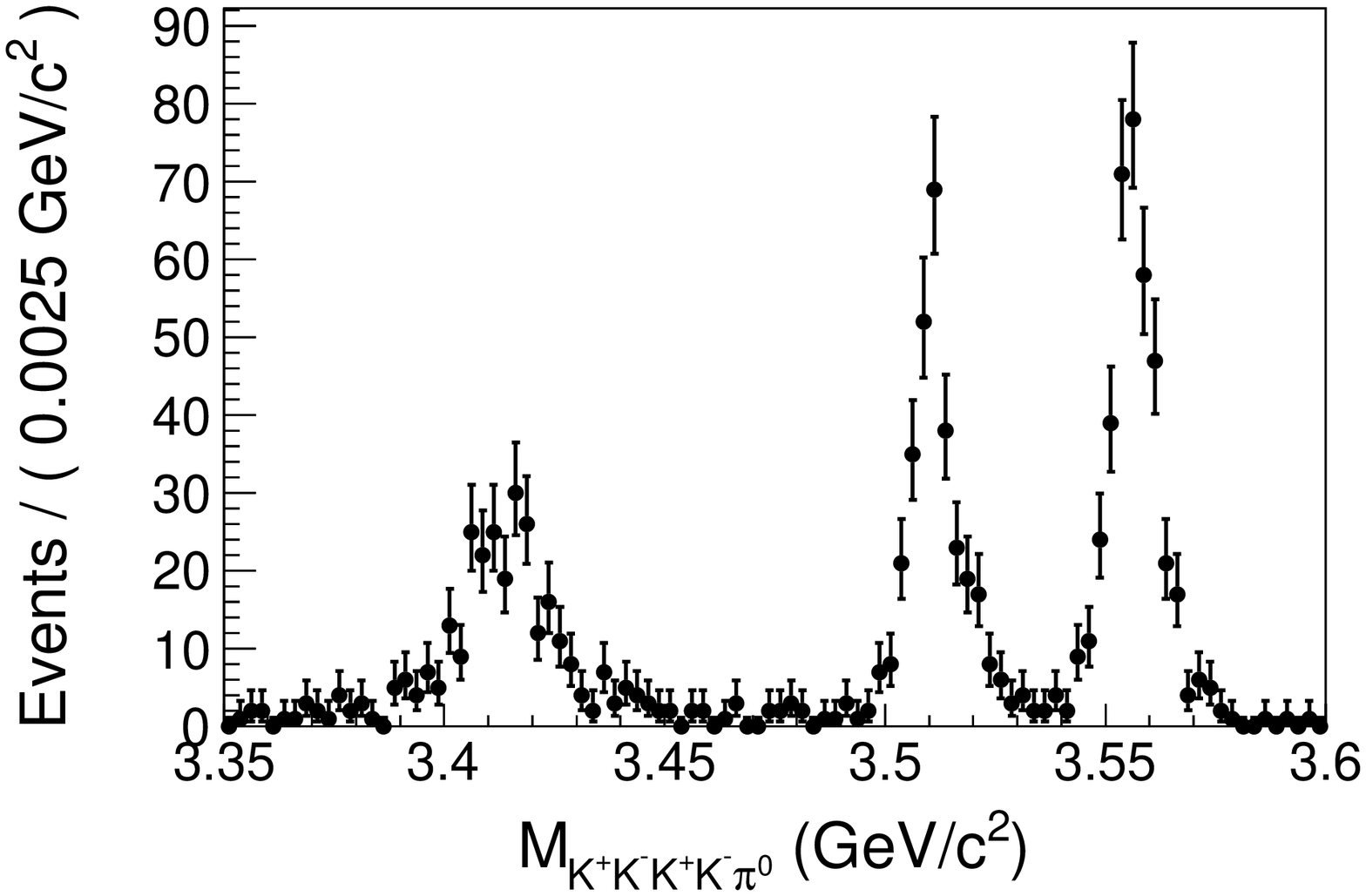}
    \put(25,50){(b)}
    \end{overpic}
    \end{center}
    \caption{%
        Invariant mass distribution of (a) $\kk\kskp$ and (b) $\kk\kkp$ with the $\phi$ mass window requirement.
     }%
    \label{fig:chicjmass}
\end{figure}

\begin{figure}[htbp]
    \begin{center}
      \begin{overpic}[width=0.45\textwidth,angle=0]{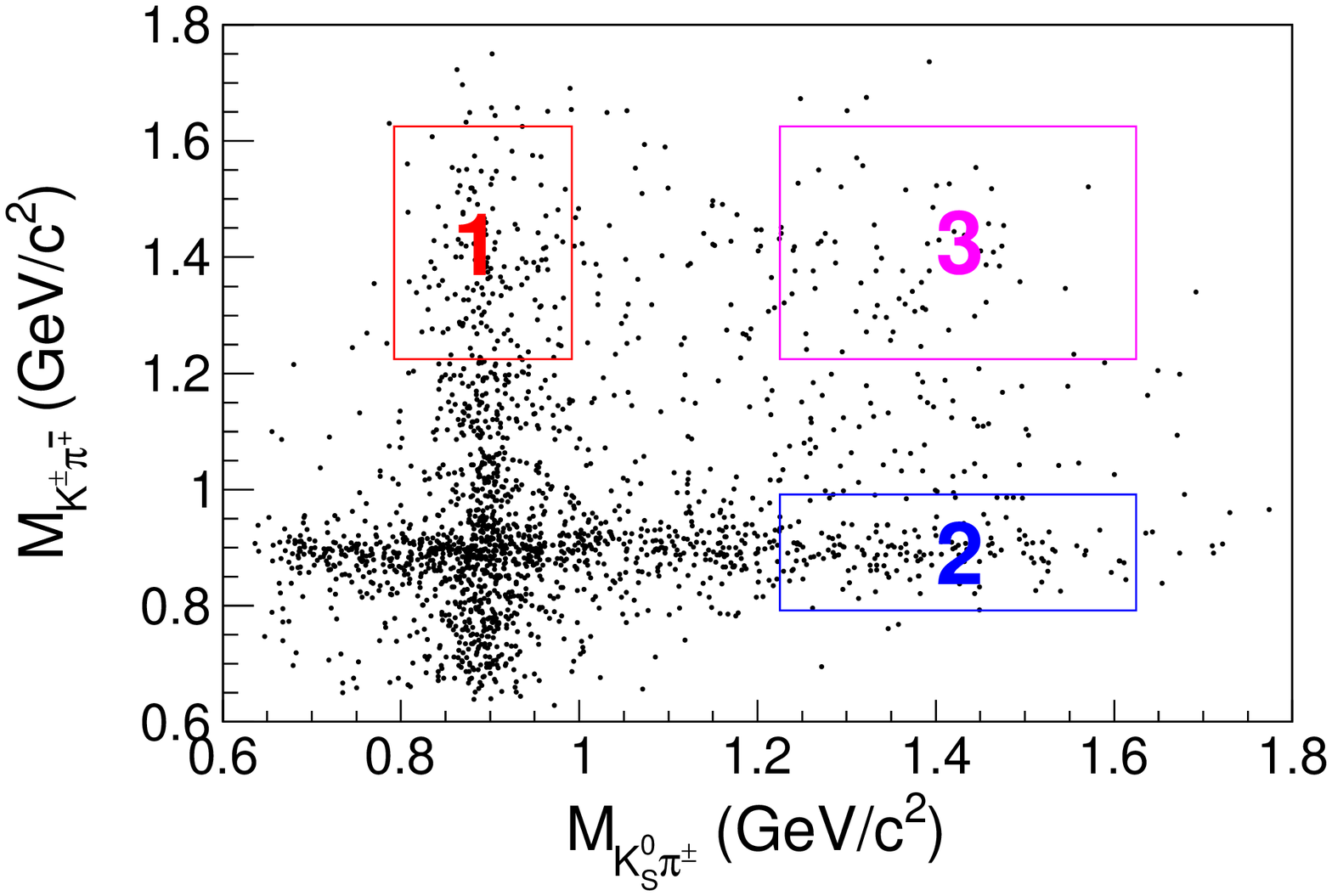}
    \put(20,50){(a)}
    \end{overpic}
    \begin{overpic}[width=0.45\textwidth,angle=0]{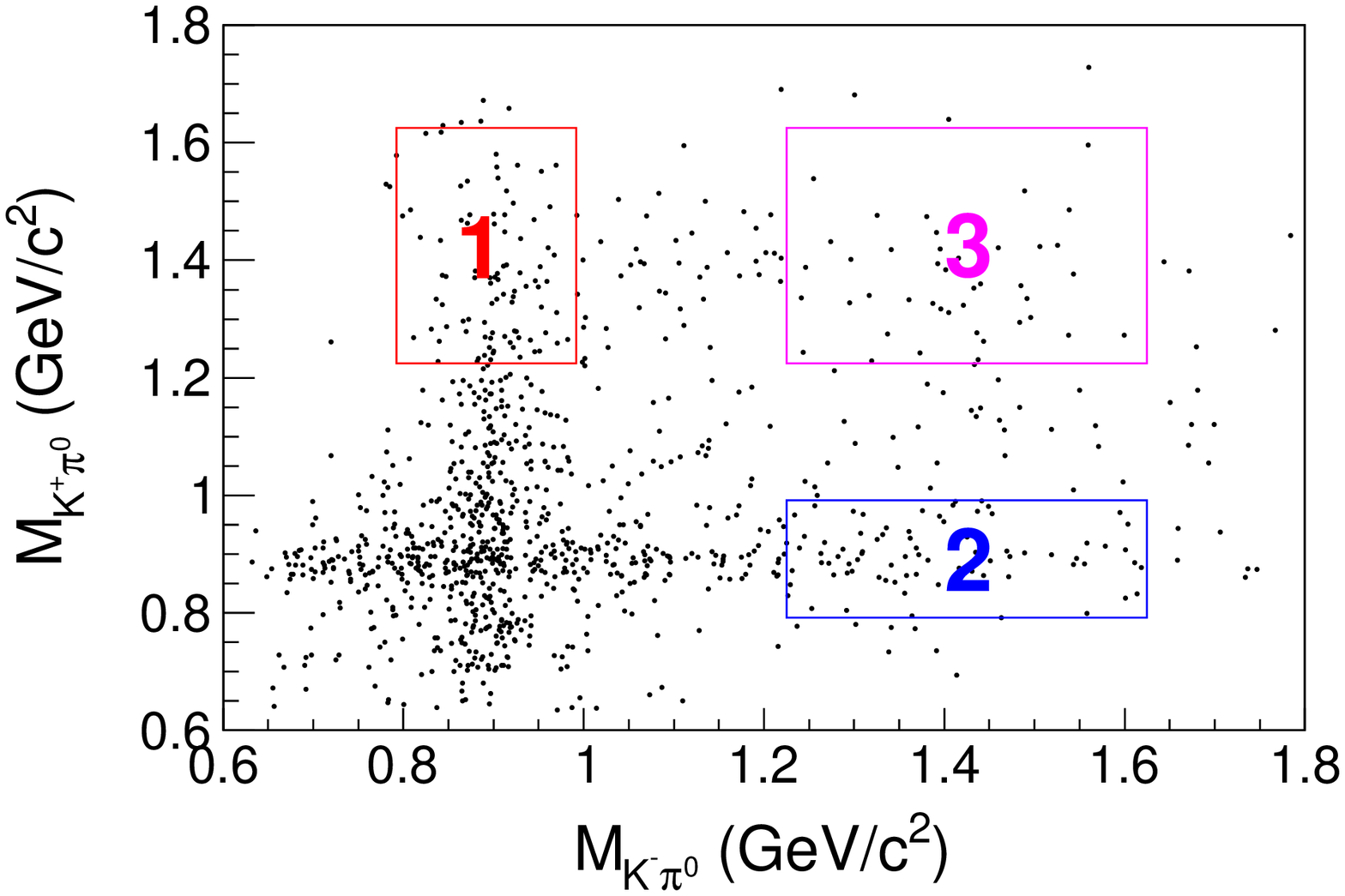}
    \put(20,50){(b)}
    \end{overpic}
    \end{center}
    \caption{%
        Scatter plot of the invariant mass of (a) $K_S^0\pi^\mp$ versus $K^\pm \pi^\mp$
        and (b) $K^+\pi^0$ versus $K^-\pi^0$ with the $\phi$ mass
        window requirement.  The boxes represent the $\kstar$ sideband
        regions described in the text.
     }%
    \label{fig:scatting}
\end{figure}

The invariant mass distributions of $\kk\kskp$ ($\kk\kkp$) with the $\phi$ mass window requirement
are shown in Fig.~\ref{fig:chicjmass}.
Significant $\ccj$ signals are observed with low background.
The corresponding scatter plots of the invariant masses of $\ksp$ versus $\kpi$
($K^+\pi^0$ versus $K^-\pi^0$) are shown in Fig.~\ref{fig:scatting}.
The dominant processes are the $\ccj\to\phi\kstark$ three body decays.

\section{\label{sec:background}Background analysis}
Since the candidate events are dominated by the three-body decay $\ccj\to\phi\kstark$,
the corresponding branching fractions are measured by imposing a mass window requirement
on the $K\pi$ invariant mass, {\it i.e.} $|M_{K\pi}-M_{\kstar}|<100$ $\mevcc$.
For convenience, hereafter, an event with $\ccj\to\phi\kstarkc$ or $\ccj\to\phi\kstarkn$
decay is called a $\kstarc$ or $\kstarn$ event.
Also, an event satisfying the requirement $|M_{\ksp}-M_{\kstarc}|<100$ $\mevcc$ or $|M_{\kpi}-M_{\kstarn}|<100$ $\mevcc$
is called a $\kstarc$ or $\kstarn$ tagged event.
The potential backgrounds for the decay $\ccj\to\phi\kstark$ are studied based on the inclusive MC sample.
The following background categories are considered:

The first background category contains the non-$\ccj$ backgrounds, which are dominated by processes
such as
$\psp\to\phi\kstar\bar{K}^{*}(892)$ and
$\psp\to\gamma\phi\kstark$.
MC studies show that these backgrounds do not produce peaks in the $\ccj$ mass region,
and their invariant mass spectrum can be described
in the fit with a polynomial function when extracting the $\ccj$
signal.

The second background category is the non-$\phi$ background, which peaks in $\ccj$ mass region.
The main processes of this background are $\ccj\to f_{0/2}^\prime \kstark$, which have
the same final state as that of the signal, where $f_{0/2}^\prime$ is a scalar or tensor
meson, {\it e.g.} $f_0(1710)$, decaying to the $\kk$ final state.
An unbinned maximum likelihood fit is performed to the $\kk$ invariant mass.
The fit result shows that the non-$\phi$ background is less than 1.8\% in the $\phi$ mass window region.
When we calculate the number of $\ccj$ events, a $\phi$ sideband
as indicated in Fig.~\ref{fig:phimass} is used to estimate the background
from non-$\phi$ events. The number of non-$\phi$ background events is subtracted to obtain
the yields of $\ccj$ signals.

The third background category is composed of non-$\kstar$ events.
From the scatter plots of the invariant mass of $\ks\pi^\mp$ versus $\kpi$ ($\kp\piz$ versus $\km\piz$)
(see Fig.~\ref{fig:scatting}), it appears as though the dominant backgrounds are the reactions containing a high mass
$K^*$ state, {\it e.g.} $\ccj\to\phi K_{0/2}^{*0}(1430)\bar{K^0}$ and $\ccj\to\phi K_{0/2}^{*\pm}(1430)K^\mp$.
This background is treated similarly to the second background category:
these processes have the same final state
as the signal and produce peaking backgrounds in the $\ccj$ mass region (background level ~1.6\%).
Like the signal, $\ccj\to\phi\kstark$,
this background category has horizontal and vertical
bands around 1.430 $\gevcc$, and contaminates in the
$\ccj\to\phi\kstark$,
illustrated as boxes 1 and 2 in Fig.~\ref{fig:scatting}.
The degree of contamination can be evaluated using the $K^*$ 2-dimensional (2D) sideband region
with $|M_{K\pi}-1425|<200~\mevcc$, illustrated as region 3 in Fig.~\ref{fig:scatting}.
The $\kk$ invariant mass for events within the $K^*$ 2D sideband region (without the $\phi$
mass window requirement) is studied.
The $\kk$ invariant mass peaks around the $\phi$ mass region,
and the non-$\phi$ events in this region are neglected in the fit.

The last background category is the QED background, which is not produced via the $\psp$ resonance.
Backgrounds of this type are evaluated using the data collected at 3.65 $\gev$ and are found to be small
and distributed uniformly in the $\ccj$ mass region.
In the fit, the contribution from this type of background is
taken into account by the polynomial function for the first background category.

\section{\label{sec:phiKS0Kpi:bkg}Signal extraction}
\begin{figure*}[htbp]
    \begin{center}
    \begin{overpic}[width=0.45\textwidth,angle=0]{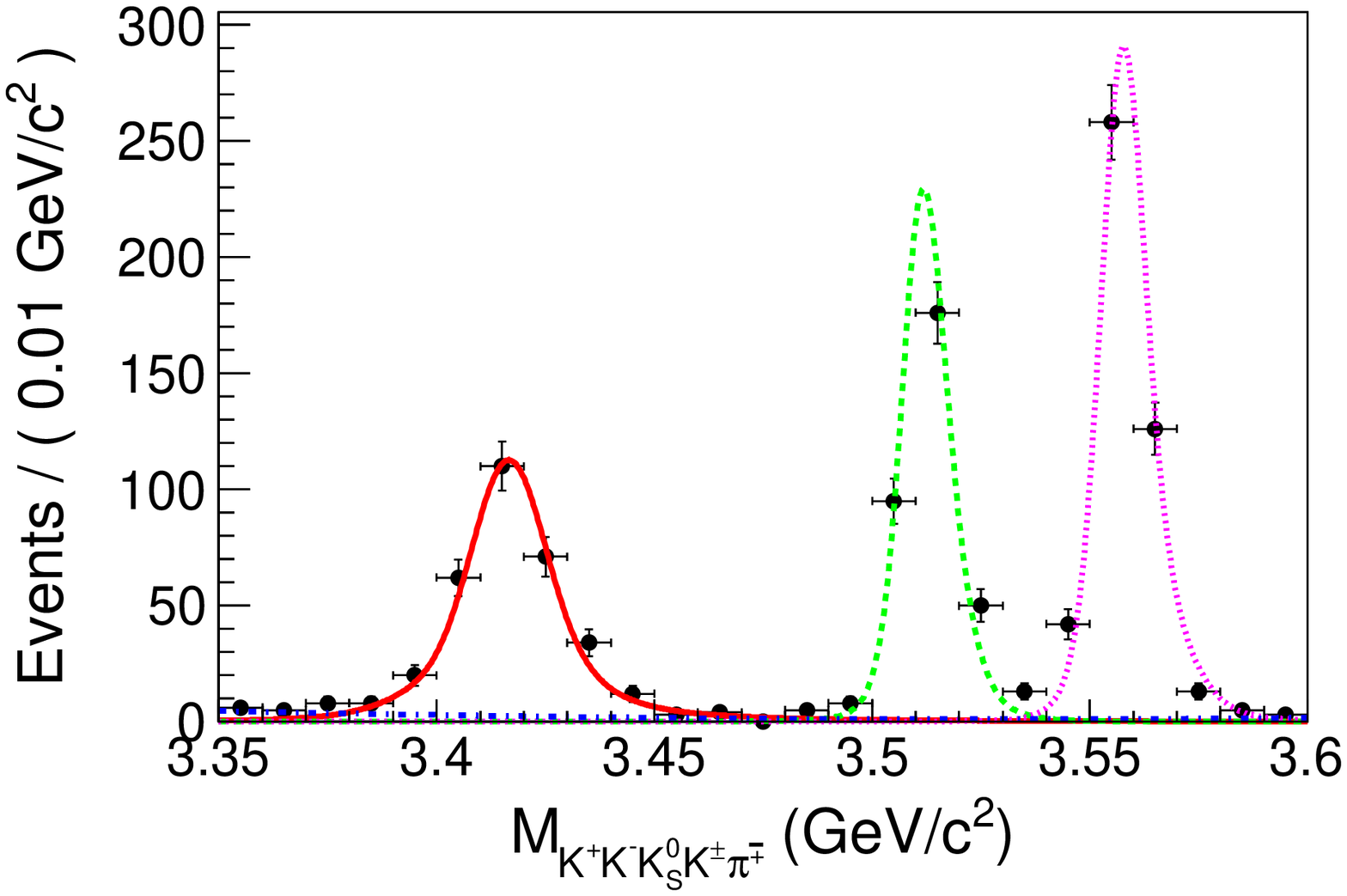}
    \put(20,50){(a)}
    \end{overpic}
    \begin{overpic}[width=0.45\textwidth,angle=0]{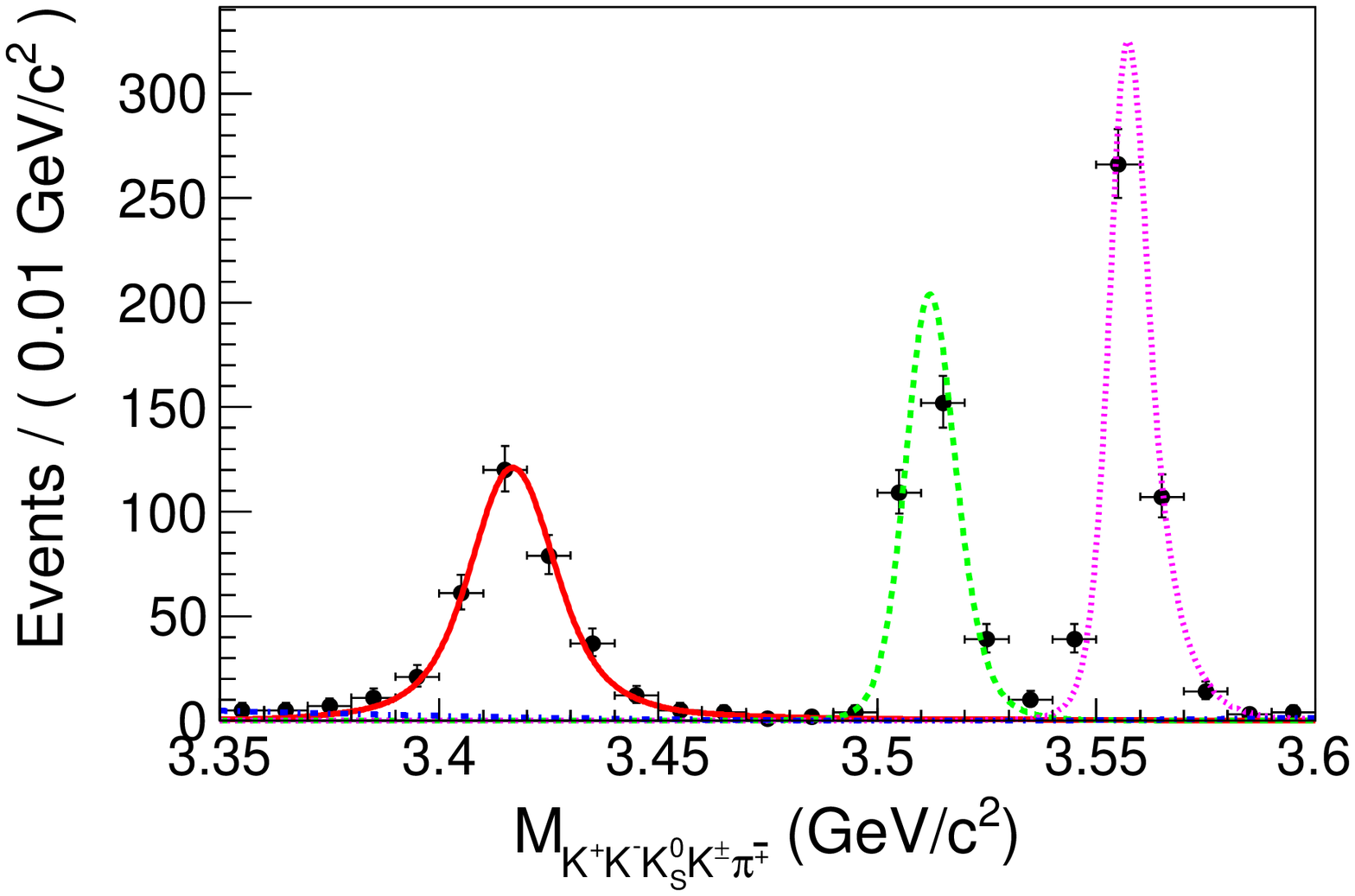}
    \put(20,50){(b)}
    \end{overpic}

    \begin{overpic}[width=0.45\textwidth,angle=0]{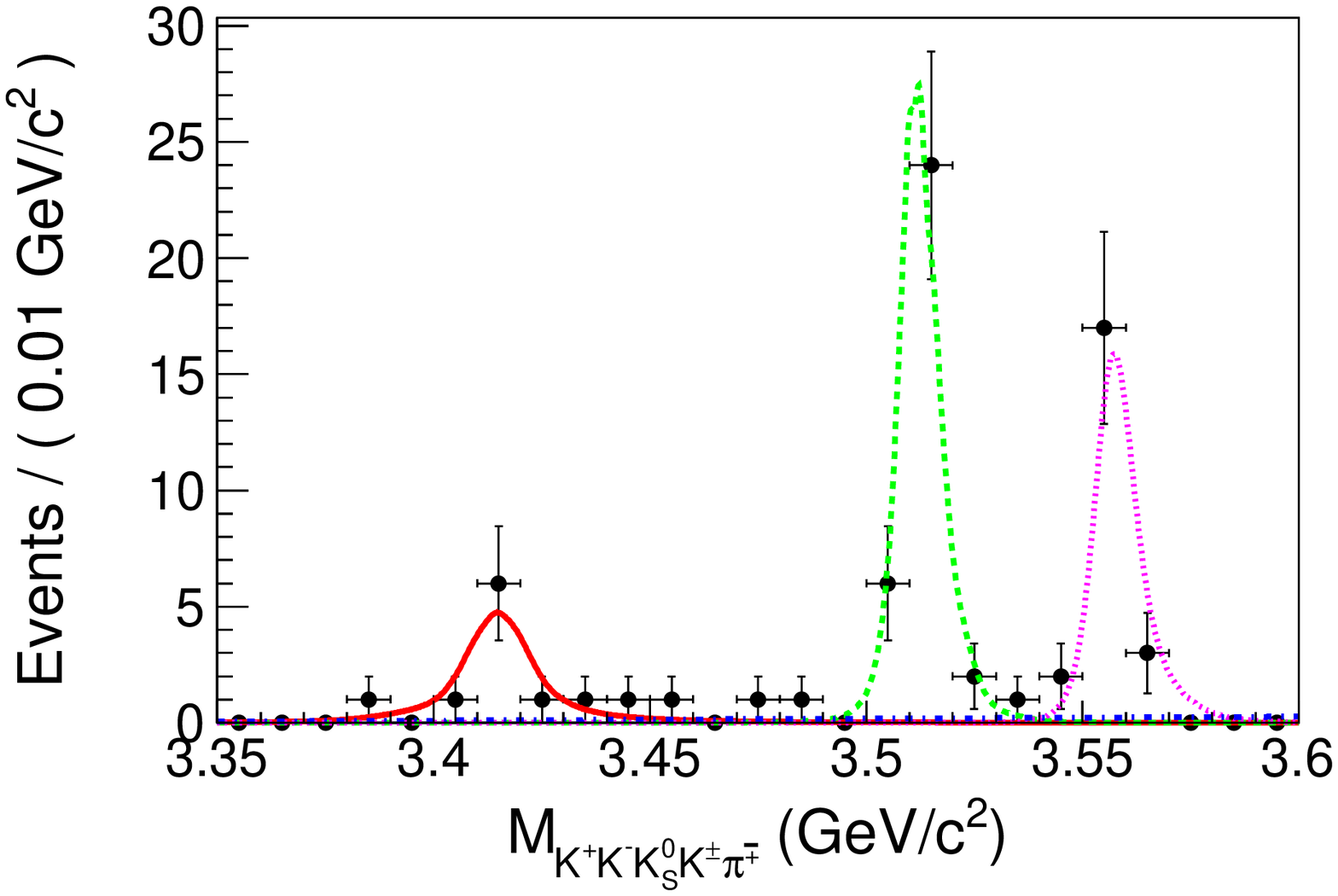}
    \put(20,50){(c)}
    \end{overpic}
    \begin{overpic}[width=0.45\textwidth,angle=0]{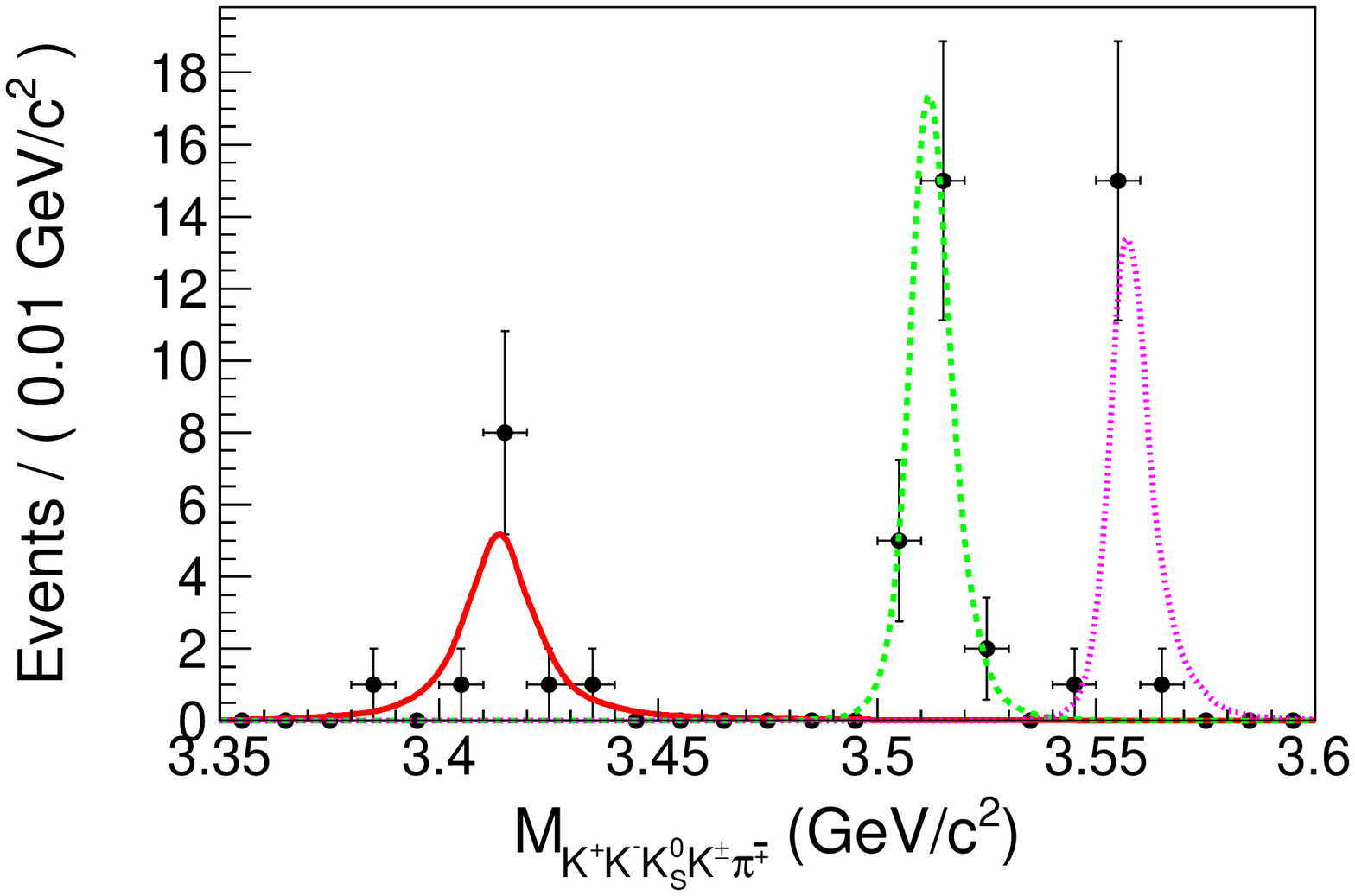}
    \put(20,50){(d)}
    \end{overpic}
    \begin{overpic}[width=0.45\textwidth,angle=0]{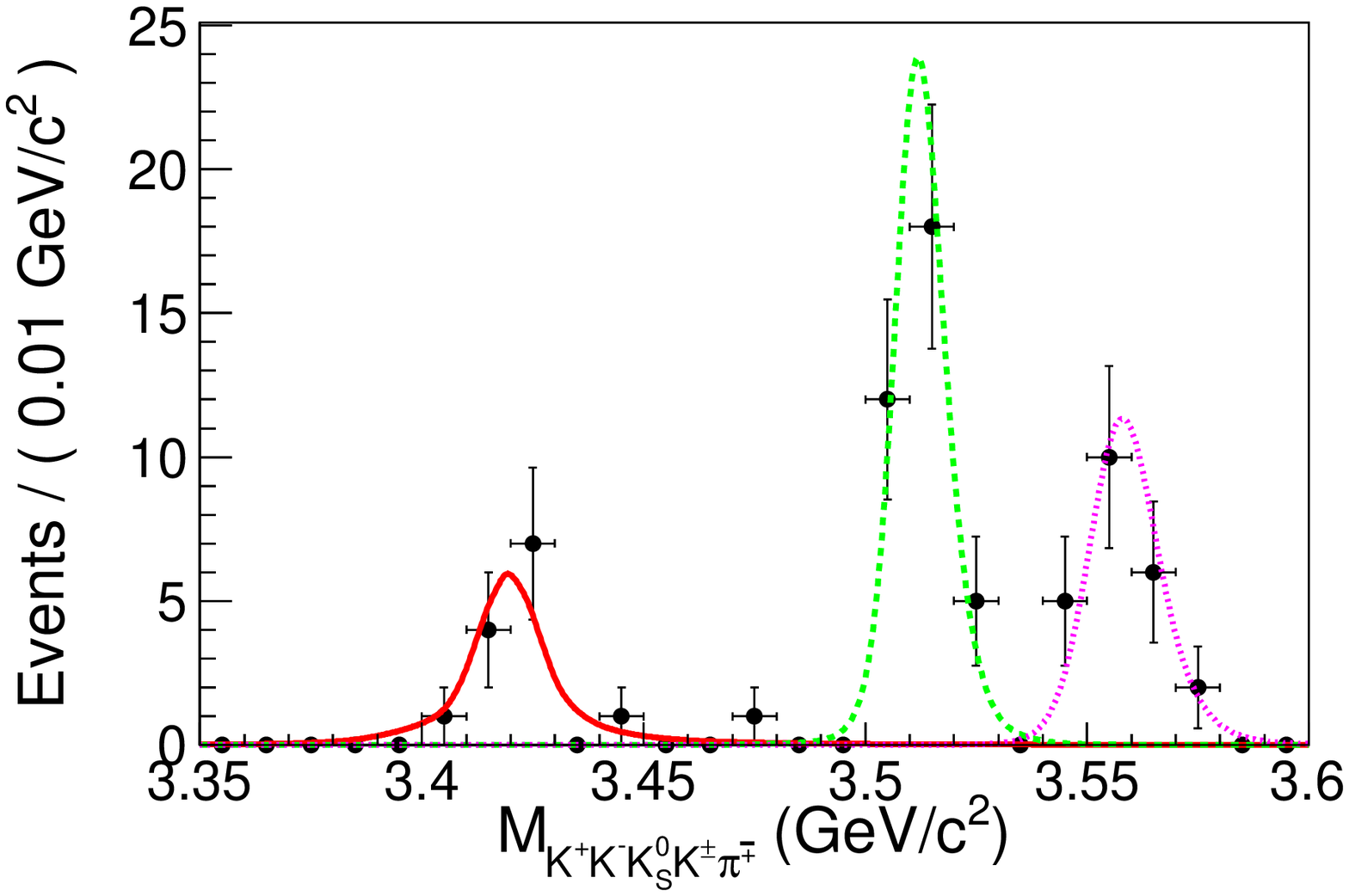}
    \put(20,50){(e)}
    \end{overpic}
    \end{center}
    \caption{%
        The invariant mass of $\kk\kskp$ distributions:
        (a) for $\kstarc$ tagged events;
        (b) for $\kstarn$ tagged events;
        (c) for $\kstarc$ tagged events in $\phi$ sideband region;
        (d) for $\kstarn$ tagged events in $\phi$ sideband region;
        (e)  for the events in the $K^*$ 2-D sideband region.
     }%
    \label{fig:chn1sg}
\end{figure*}

For the $\ccj\to\phi\kskp$ decay, the isospin conjugate channels $\ccj\to\phi
\kstarkn$ and $\ccj\to\phi\kstarkc$ are included with $\kpi$ forming a $\kstarn$ or with $\ksp$
forming a $\kstarc$. The branching fractions of these reactions are measured separately.
The invariant mass of the $\kk\kskp$ in different regions is shown in Fig.~\ref{fig:chn1sg} for (a) $\kstarc$ tagged events,
(b) $\kstarn$ tagged events, (c) $\kstarc$ events
within the $\phi$ sideband region, (d) $\kstarn$ events
within the $\phi$ sideband region, (e) events in the $K^*$
2D-sideband region.

Unbinned maximum likelihood fits are performed to the invariant mass spectra of $\kk\kskp$ to extract
the yields of $\ccj$ signal in different regions.
In the fits, the three $\ccj$ signals are described by the corresponding MC shapes convoluted with Gaussian
functions which represent the difference in resolution between the data and MC. The widths of the Gaussian functions are float.
The background is represented by a second order Chebyshev polynomial function.
The results of the fits are shown in Fig.~\ref{fig:chn1sg}, and the
signal yields are summarized in Table~\ref{tab:chn1entry}, where
$N^{\sig}_\mathrm{obs}(K^{*\pm})$ and $N^{\sig}_\mathrm{obs}(K^{*0})$ are the number
of $\kstarc$ and $\kstarn$ tagged events,
$N_{\phi~\mathrm{sideband}}^{\bkg}(K^{*\pm})$ and
$N_{\phi~\mathrm{sideband}}^{\bkg}(K^{*0})$ are those in the $\phi$ sideband
regions, and $N^{\bkg}_{K^{*}~\mathrm{sideband}}$ in the $K^*$ 2-D sideband region.

\begin{table*}[htbp]
\caption{The $\ccj$ yields in different regions from fitting the $\kk\kskp$ final state.
         The uncertainties shown are statistical only.}
\label{tab:chn1entry}
\begin{center}
\begin{tabular}{cccccc}
\hline
 &~~~$N^{\sig}_\mathrm{obs}(K^{*\pm})$~~&~~$N_{\phi~\mathrm{sideband}}^{\bkg}(K^{*\pm})$~~&~~$N^{\sig}_\mathrm{obs}(K^{*0})$~~&~~ $N_{\phi~\mathrm{sideband}}^{\bkg}(K^{*0})$~~&~~$N_{K^*~\mathrm{sideband}}^{\bkg}$~~    \\
\hline
\hline
$\chi_{c0}$   & $317\pm21$  & $11\pm5$ & $349\pm21$  & $12\pm3$ &  $14\pm4$  \\
$\chi_{c1}$   & $329\pm19$  & $33\pm6$ & $310\pm18$  & $22\pm5$ &  $35\pm6$  \\
$\chi_{c2}$   & $443\pm22$  & $21\pm5$ & $428\pm21$  & $17\pm4$ &  $23\pm5$  \\
\hline
\end{tabular}
\end{center}
\end{table*}

\begin{figure}[htbp]
    \begin{center}
    \begin{overpic}[width=0.45\textwidth,angle=0]{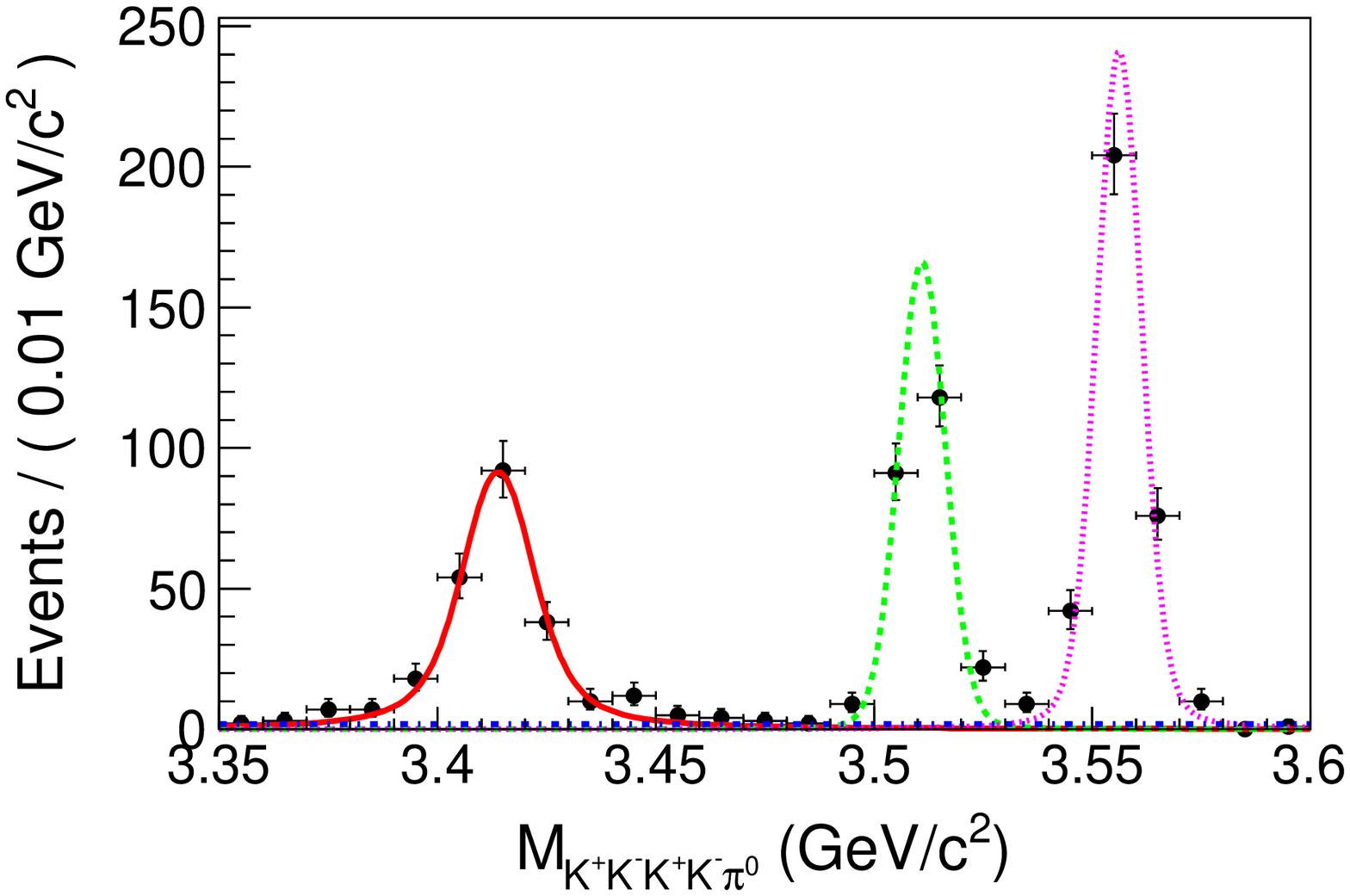}
    \put(20,50){(a)}
    \end{overpic}
    \begin{overpic}[width=0.45\textwidth,angle=0]{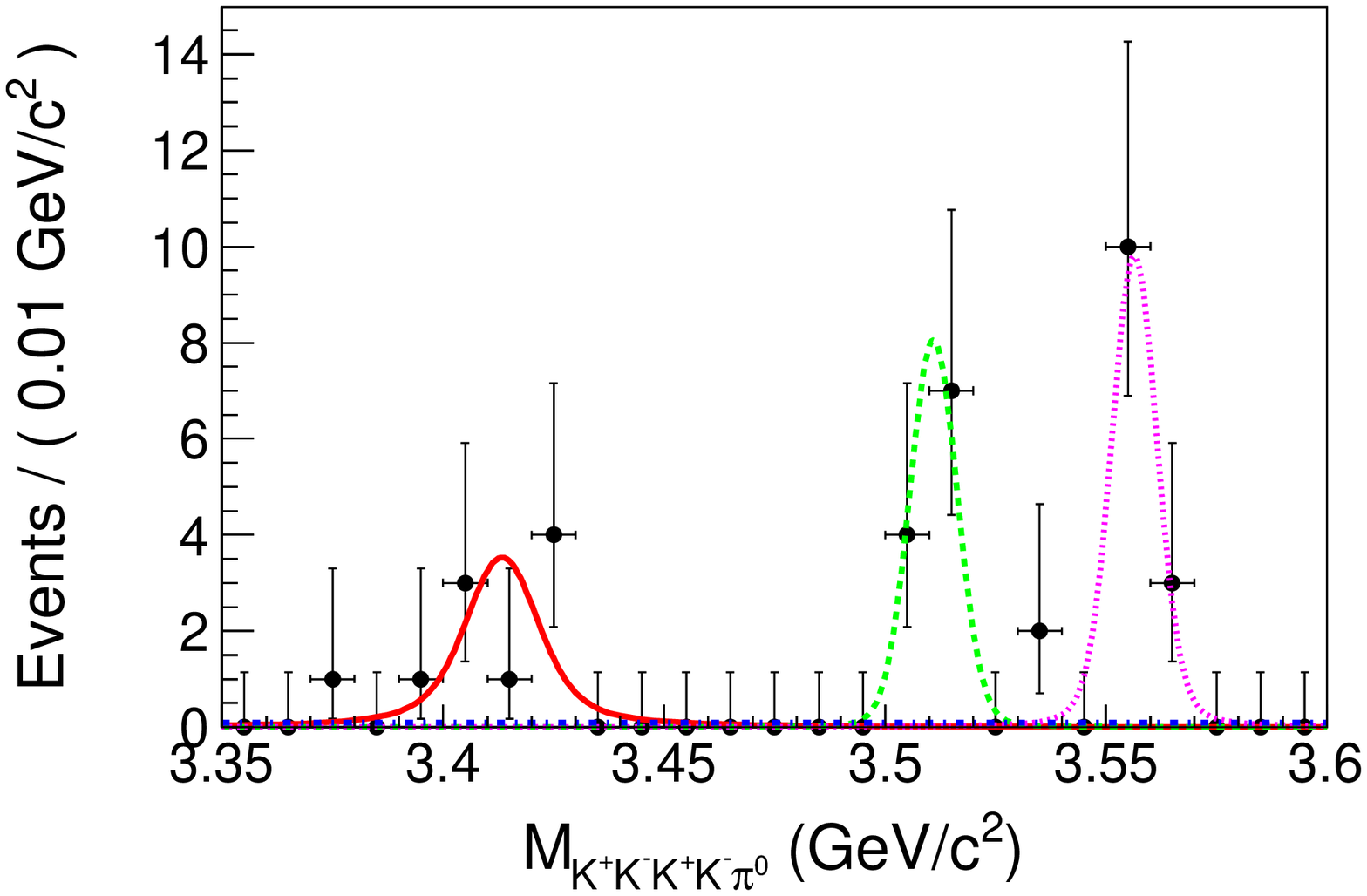}
    \put(20,50){(b)}
    \end{overpic}
    \begin{overpic}[width=0.45\textwidth,angle=0]{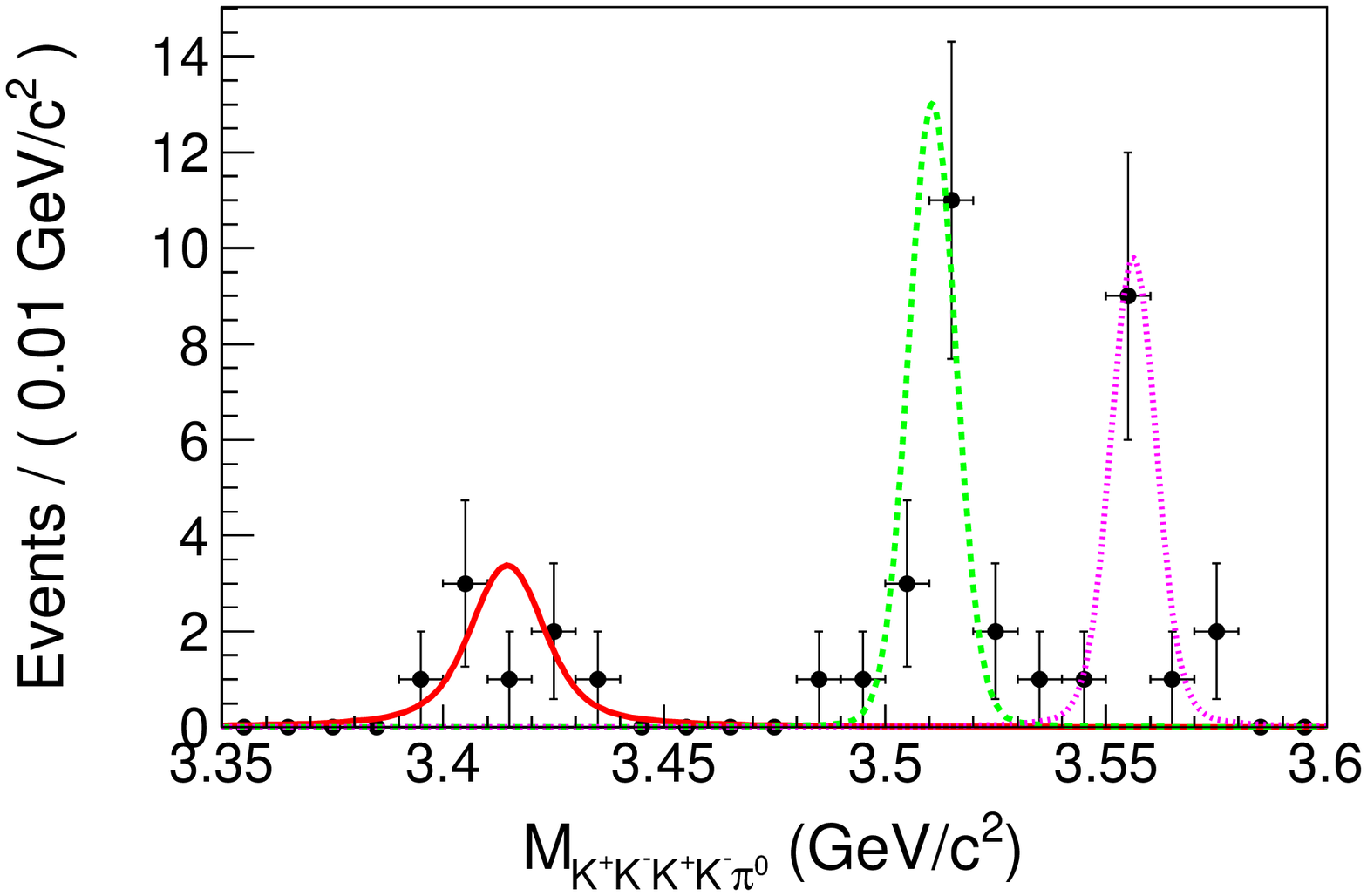}
    \put(20,50){(c)}
    \end{overpic}
    \end{center}
    \caption{
        The invariant mass distributions of $\kk\kkp$ for the candidate events
        (a) within the signal region, (b) within the $\phi$ sideband region,
        (c) within the $K^*$ 2-D sideband region.
     }
    \label{fig:chn2sg}
\end{figure}

For the $\ccj\to\phi\kkp$ decay mode, only the charged $\kstarc$ state is included.
In this analysis, the charge conjugate modes are not separated, and the joint branching
fraction of $\ccj\to\phi\kstarkc$ is measured.
Figure~\ref{fig:chn2sg} shows the distributions of the $\kk\kkp$
invariant mass for the signal, $\phi$ sideband, and $K^*$ 2-D sideband regions.

The same fits are performed to the individual distributions. The fit curves are
shown in Fig.~\ref{fig:chn2sg}, and the resulting $\ccj$ signal yields are summarized
in Table~\ref{table:fitchicj-kkpi0-data}, where $N^{\sig}_\mathrm{obs}(K^{*\pm})$ is the number of events in
the signal region, and $N_{\phi~\mathrm{sideband}}^{\bkg}(K^{*\pm})$
and $N_{K^*~\mathrm{sideband}}^{\bkg}$ are those in the $\phi$ and
$K^*$ 2-D sideband region. Here, $\epsilon$ is the detection efficiency.

\begin{table*}[htbp]
\caption{ The $\ccj$ yields in different regions and the
          detection efficiency from MC samples with $\kk\kkp$ final state.
          The uncertainties are statistical only.
          The statistical errors on the efficiencies are neglected
          because of the large size of the MC samples used.
         \label{table:fitchicj-kkpi0-data}}
\begin{center}
\begin{tabular}{ccccc}
\hline
                & ~~~$N^{\sig}_\mathrm{obs}(K^{*\pm})$~~~& ~~~$N_{\phi~\mathrm{sideband}}^{\bkg}(K^{*\pm})$~~~&~~~$N_{K^*~\mathrm{sideband}}^{\bkg}$~~~&~~~~~$\epsilon$~~~~~ \\
\hline
\hline
~~~$\chi_{c0}$~~~& $231\pm17$  &  $9\pm3$    & $9\pm3$  &   6.93\%  \\
~~~$\chi_{c1}$~~~& $229\pm16$  & $11\pm3$   & $18\pm4$  &   8.23\%  \\
~~~$\chi_{c2}$~~~& $332\pm19$  & $14\pm4$   & $14\pm4$  &   7.56\%  \\
\hline
\end{tabular}
\end{center}
\end{table*}

\section{Branching fractions}

The branching fractions are calculated using the extracted signal yields that have been corrected according to the
detector efficiency and decay branching fractions of intermediate states.

For the $\ccj\to\phi \kskp $ decay mode, a clear intersection is observed on the scatter plot
of the invariant mass of $\ksp$ versus $\kpi$ (Fig.~\ref{fig:scatting}~(a)).
This indicates that some $\kstarc$ events are included in the $\kstarn$ tagged events,
and vice versa.
Without consideration of the interference between two isospin conjugated $K^{*}(892)$s in $\chi_{cJ} \to \phi K^{\pm} K_{S}^{0} \pi^{\mp}$ and charge conjugated $K^{*}(892)$s in $\chi_{cJ} \to \phi K^{+} K^{-} \pi^{0}$, the relationship between the observed signal yields and the corresponding
branching fractions is given by the following equations:
\begin{eqnarray}
N_{c}^{\sig} = N_{\psp}Br_c\epsilon_{cc}+N_{\psp}Br_n\epsilon_{nc}, \\
N_{n}^{\sig} = N_{\psp}Br_c\epsilon_{cn}+N_{\psp}Br_n\epsilon_{nn},
\end{eqnarray}
where $N_{\psp}$ is the number of $\psp$ events,
$N_{c/n}^{\sig}$ is the number of $K^{*}(892)^{\pm/0}$ tagged candidates,
and $Br_{c/n}$ is the product branching fraction  $Br(\psp\to\gamma\ccj) \times Br(\ccj \to \phi \kstarkc / \phi \kstarkn)  \times
Br(\phi\to\kk)  \times Br(\kstarc\to K^{0}\pi^{\pm} / \kstarn\to\kpi)\times Br(\bar{K^{0}} / K^{0} \to K^{0}_S) \times Br(\ks\to \pp)$  where $Br(\psp\to\gamma\ccj)$, $Br(\phi\to\kk)$, $Br(\kstarc\to K^{0}\pi^{\pm} )$, $Br(\bar{K^{0}} / K^{0} \to K^{0}_S)$, $Br(\kstarn\to\kpi)$ and
$Br(\ks\to \pp)$ are taken from the PDG~\cite{pdg},
and the branching fractions of $\ccj\to\phi\kstarkc$ / $\phi\kstarkn$
are to be measured in this analysis.
The factor $\epsilon_{cc/nn}$ is the detection efficiency for $K^{*}(892)^{\pm/0}$ events to be
identified as $K^{*}(892)^{\pm/0}$ tagged events,
and $\epsilon_{cn/nc}$ is the detection efficiency for $K^{*}(892)^{\pm/0}$ events to be identified as
$K^{*}(892)^{\pm/0}$ tagged events.
Then the branching fractions are calculated as :
\begin{align}
\label{calbrAB}
{\it Br}_c &= \frac{N_c^{\sig}\epsilon_{nn}-N_n^{\sig}\epsilon_{nc}}
{N_{\psp}(\epsilon_{cc}\epsilon_{nn}-\epsilon_{nc}\epsilon_{cn})} \\
{\it Br}_n &= \frac{N_c^{\sig}\epsilon_{cn}-N_n^{\sig}\epsilon_{cc}}
{N_{\psp}(\epsilon_{cn}\epsilon_{nc}-\epsilon_{nn}\epsilon_{cc})}.
\end{align}
The corresponding background subtracted signal yields $N^{\sig}_{c/n}$
are calculated as:
\begin{align}
N^{\sig}_{c/n} &= N^{\sig}_\mathrm{obs}(K^{*\pm/0}) \nonumber     \\
&\quad - ~f_{\phi~\mathrm{sideband}} ~N^{\bkg}_{\phi~\mathrm{sideband}}(K^{*\pm/0}) \\
&\quad - ~f_{K^*~\mathrm{sideband}} ~N^{\bkg}_{K^*~\mathrm{sideband}}. \nonumber
\end{align}
Here, $f_{\phi~\mathrm{sideband}}$ and $f_{K^*~\mathrm{sideband}}$ are normalization
factors; based on the $\kk$ invariant mass distribution in
Fig.~\ref{fig:phimass}(a), $f_{\phi~\mathrm{sideband}}$ is taken as 0.37, while
$f_{K^*~\mathrm{sideband}}$ is taken to be 1/4. This takes into account the area of the sideband region
(box 3 in Fig.~\ref{fig:scatting}) relative to that of polluted signal regions (box 1 or 2 in Fig.~\ref{fig:scatting})
as well as the existence of both isospin conjugate channels in the sideband regions.
The detection efficiencies are evaluated with exclusive signal MC samples, which contain events in the reactions
$\psp\to\gamma\ccj$, $\ccj\to\phi\kstarkc/\phi\kstarkn$ with subsequent decay
$\phi\to\kk$, $\kstarc\to K^{0}\pi^{\pm}$ or $\kstarn\to\kpi$, $\bar{K^{0}} / K^{0} \to \ks$, $\ks\to\pp$.
The $\psp \to \gamma \ccj$ decay is generated with a $1 + \lambda \cos^{2}\theta$ distribution,
where $\theta$ is the angle between the direction of the radiative photon and that of the positron beam,
and $\lambda$ = 1, -1/3, 1/13 for $J$ = 0, 1, 2, assuming pure E1 transitions.
The $\ccj$ decays are generated with a flat angular distribution.
Due to the existence of resonances in the $\kstarkc$ invariant mass,
the detection efficiency is evaluated by weighting the signal MC samples by the $\kskp$ invariant mass. This accounts for
differences in the distributions between the data and the MC simulation.
The detection efficiencies are listed in Table~\ref{tab:phiKS0Kpi:effs}
and the corresponding measured branching fractions $Br(\ccj\to\phi\kstarkc/\phi\kstarkn)$
are given in Table~\ref{table:brfractionkskpisum}.

\begin{table}[htbp]
\caption{Detection efficiencies from MC samples. The errors of efficiencies are neglected because of large MC samples.}
\label{tab:phiKS0Kpi:effs}
\begin{center}
\begin{tabular}{ccccc} \hline
            &~~~~~$\epsilon_{cc}$~~~~~&~~~~~$\epsilon_{cn}$~~~~~&~~~~~$\epsilon_{nc}$~~~~~&~~~~~$\epsilon_{nn}$~~~~~ \\
\hline \hline
$\chi_{c0}$~~~&   $10.47\%$     &   $4.08\%$      &    $4.29\%$      &   $10.80\%$    \\
$\chi_{c1}$~~~&   $11.43\%$     &   $4.34\%$      &    $4.58\%$      &   $11.96\%$   \\
$\chi_{c2}$~~~&   $11.42\%$     &   $4.75\%$      &    $4.49\%$      &   $11.48\%$    \\
\hline
\end{tabular}
\end{center}
\end{table}

For the $\ccj\to\phi\kkp$ decay mode, only the charged $\kstarc$ intermediate state is included,
and the joint branching fractions of the charged conjugate modes are calculated by
\begin{eqnarray}
Br(\ccj\to\phi \kstarkc) =  \nonumber \\          \frac{N_\mathrm{obs}^{\sig} - 0.25 * N_{\phi}^{\bkg} - 0.5 * N_{K^*}^{\bkg}}
{\epsilon \cdot Br \cdot N_{\psp} }.
\end{eqnarray} where $Br$ is the product branching fraction of the
other processes in the cascade decay
including $\psp\to\gamma\ccj$, $\phi\to\kk$, $\kstarc\to K^\pm\piz$ and $\pi^0\to\GG$,
and all the individual branching fractions are taken from PDG.
A factor of 0.25 is determined by the $\kk$ invariant mass distribution in Fig.~\ref{fig:phimass}(b).
The $K^*$ 2-D background is taken with a weight of 0.5 due to the double area of box 3 in
Fig.~\ref{fig:scatting} (b), compared to box 1 or 2.
As described above, the detection efficiencies are evaluated using the exclusive MC samples weighted by
the $\kkp$ invariant mass.

\section{Systematic uncertainties}

Several sources of systematic uncertainties are considered in the measurement of the branching fractions.
These include differences between the data and the MC simulation for track reconstruction, PID, photon detection,
kinematic fitting, $\piz$ selection, $\ks$ reconstruction, the mass window requirement, the fitting process, background estimation,
MC modeling, the branching fractions of intermediate state decays, and the luminosity measurement.

{\it (a) Tracking and PID efficiency.~}
The tracking efficiencies for $K^{\pm}$ and $\pi^{\pm}$ as functions of transverse momentum have been
investigated with the control samples $\jpsi\to\kskp,\ks\to\pp$~\cite{kaoneffi} and $\psp\to\pp\jpsi$~\cite{pioneffi}.
The uncertainty of the tracking efficiency is $1\%$ for each pion and
$1.58\%$ for each kaon.
These uncertainties are obtained taking the
transverse momentum distributions found in data into account.

The uncertainty due to PID has been studied via the same control
samples~\cite{pioneffi} and is estimated to be $2\%$  for each charged
pion and $2.23\%$ for each charged kaon by weighting according to the
transverse momentum distributions.

{\it (b) Photon detection efficiency.~}
The uncertainty due to photon detection efficiency is $1\%$ per photon~\cite{photoneffi}.
This is determined from studies of photon detection efficiencies with a large and high purity control
sample of $\jpsi\to\rho^{0}\piz$ where $\rho^{0}\to\pp$ and $\piz\to\gamma\gamma$.

{\it (c) $4C$-kinematic fit.~}
In the analysis, the track helix parameters ($\phi_{0}$, $\kappa$, $\tan\lambda$)
are corrected for the MC sample in order to reduce the difference of the $4C$ kinematic fit $\chi^{2}_{\rm 4C}$
between the data and the MC sample. Here, $\phi_{0}$ is the azimuthal angle that specifies the pivot with
respect to the helix center, $\kappa$ is the reciprocal of the transverse momentum
and $\tan\lambda$ is the slope of the track.
The correction factors are obtained from a clean sample of $\jpsi\to\phi f_{0}(980)$, $\phi\to\kk$
and $f_{0}(980)\to\pp$.
An alternative detection efficiency is evaluated with the same MC samples, but without helix
parameters corrections. The difference in the efficiencies is taken as the
uncertainty of the $4C$ kinematic fit~\cite{guoyp}.

{\it (d) $\piz$ selection.~}
The uncertainty due to the $\piz$ selection is determined from a high purity control sample of $\jpsi\to\pp\piz$.
The difference in the $\piz$ selection efficiency between the data and the MC simulation, 1.0\%, is taken as the uncertainty
for the $\piz$ selection~\cite{photoneffi}.

{\it (e) $\ks$ reconstruction.~}
The uncertainty for the $\ks$ reconstruction efficiency is studied with a control sample of $\jpsi\to
\kstarkc$. A conservative value of 3.5\% is taken as the systematic uncertainty for $\ks$ reconstruction~\cite{kserr}.
This uncertainty is dominated by two sources: one is the tracking efficiency of the two pions
from the $\ks$ decay, and the other is the secondary vertex fit for the two pions and the related
selection criteria.

{\it (f) $\phi$ and $\kstar$ mass window requirement.~}
The uncertainty from the $\phi$ mass window requirement is estimated by changing the $\phi$ signal
and sideband windows by one time the mass resolution
of $\phi$, $5 \mev$, which is determined by a fit to data in Fig.~\ref{fig:phimass} . The difference in the branching
fractions is taken as the uncertainty.
For the uncertainty related to the $\kstar$ mass window requirement, the value
for the $\kstar$ width used in the MC simulation, $\Gamma =
50.8\pm0.18~\mev$ \cite{pdg}, is changed by twice the uncertainty
quoted by PDG,
and the difference in the detection efficiency is taken as the systematic uncertainty.

{\it (g) Fitting process.~}
To estimate the uncertainties from the fitting process, three effects are considered.
{\it (1) $\ccj$ signal lineshape.}
The $\ccj$ signal lineshapes are described with MC simulated lineshapes
convoluted with Gaussian functions.
An alternative fit with Breit-Wigner functions convoluted with Gaussian functions for the $\ccj$
signals is performed, where the Gaussian functions represent the mass resolution. The
difference in the production yield returned by the fits is considered as the systematic uncertainty related to the signal
lineshape.
{\it (2) background lineshape.}
The non-peaking background is described with a second order Chebyshev
polynomial. Alternative fits with different order Chebyshev polynomial
functions are performed. The largest difference in the branching fractions is taken
as the systematic uncertainty.
{\it (3) fitting range.}
The invariant mass of $\kk\kskp$ ($\kk\kkp$) is fitted in the region
of [3.35, 3.6] $\gevcc$. Alternative fits with different ranges, [3.3, 3.6] $\gevcc$
or [3.3, 3.65] $\gevcc$ are performed. The maximum difference in the branching
fractions are treated as the systematic uncertainty.

{\it (h) Peaking backgrounds.~}
The peaking backgrounds without a $\phi$ signal are estimated with the
events in the $\phi$ sideband region [1.10, 1.13] $\gevcc$. The corresponding uncertainty
is studied by changing the $\phi$ sideband to the range [1.08, 1.11] $\gevcc$.
The uncertainty of the peaking background with an excited $K^*$ is estimated
by the difference in the branching fractions with or without the $K^*$ background subtraction.

{\it (i) Weighting method in MC.~}
To obtain the detection efficiency, the MC samples are weighted
by the $\kstark$ invariant mass to compensate for the difference between the data and MC simulation.
To get the uncertainty from this weighting method, the weight in each $\kstark$ invariant
mass bin is randomly changed around its mean value by a Gaussian with
a standard deviation given by the statistical uncertainty.
The same process is performed ten thousand times, and the standard deviation
on the detection efficiencies is taken as the systematic uncertainty related with the
weighting method.

{\it (j) $K^{*}(892)$ polarization.~}
We estimate the uncertainties depending on the $K^{*}(892)$ polarization by comparing the angular distribution
of final states between data and MC. We treat the differences in efficiencies as the uncertainties.

{\it (k) Other uncertainties.~}
The uncertainty of the total number of $\psp$ decays is 0.8$\%$. This uncertainty is determined from a study
of inclusive $\psp$ hadronic decays~\cite{npsp}.
The uncertainties due to the branching fractions of intermediate states
are taken from the PDG~\cite{pdg}.

Table ~\ref{tab:phiKS0Kpi:syserr} and \ref{table::sys error} summarize the systematic
uncertainties for the $\ccj\to\phi\kskp$ and $\ccj\to\phi\kkp$ decay modes, respectively.
Assuming all of the uncertainties are independent, the total systematic uncertainties are
obtained by adding the individual contributions in quadrature.

\begin{table*}[htbp]
\caption{\label{tab:phiKS0Kpi:syserr} Systematic uncertainties on the branching fraction for the
$\chi_{cJ}\to\phi K_S^0K^\pm\pi^\mp$ final states (in \%).}
\footnotesize
\begin{center}
\begin{tabular}{l c c c }
\hline
Sys. err source                         &$\chi_{c0}$      &$\chi_{c1}$     &$\chi_{c2}$ \\
\hline
\hline
\multicolumn{4}{c}{Common contribution}\\
\hline
Total number of $\psp$  & \multicolumn{3}{c}{0.8} \\
Tracking                       & \multicolumn{3}{c}{$5.7$} \\
Particle ID                    & \multicolumn{3}{c}{$8.7$} \\
Photon selection               & \multicolumn{3}{c}{1} \\
Kinematic Fit                   & \multicolumn{3}{c}{1} \\
$K_S^0$ reconstruction         & \multicolumn{3}{c}{3.5} \\
$\phi$ mass window             & \multicolumn{3}{c}{1} \\
\hline
\multicolumn{4}{c}{$\chi_{cJ}\rightarrow\phi K^{*}(892)^{\pm}K^{\mp}$ ($\chi_{cJ}\rightarrow\phi K^{*}(892)^{0}\bar{K^0}$)}\\
\hline
$K^*$ mass window               ~~~~&  1.0 (2.1)       ~~~~& 0.5 (0.6)       ~~~~& 1.0 (2.1) \\
$\chi_{cJ}$ lineshape           ~~~~&  1.5 (3.4)       ~~~~& 2.3 (3.0)       ~~~~& 0.0 (3.7)\\
Fit range                                    ~~~~&  2.0 (4.2)       ~~~~& 1.1 (1.2)       ~~~~& 0.7 (1.4)\\
Non-peaking BG shape        ~~~~&  1.0 (0.0)       ~~~~& 0.5 (0.0)       ~~~~& 0.7 (0.5)\\
Peaking BG without $\phi$ ~~~~&  0.0 (0.5)       ~~~~& 0.6 (0.7)       ~~~~& 0.8 (0.0)\\
Peaking BG with $\phi$       ~~~~&  1.0 (0.8)       ~~~~& 3.2 (3.0)       ~~~~& 1.3 (1.4)\\
BRs from PDG                           ~~~~&  2.7 (2.7)       ~~~~& 3.3 (3.3)      ~~~~& 3.4 (3.4)\\
MC model                                 ~~~~&  2.7 (2.6)       ~~~~& 2.3 (2.8)      ~~~~& 2.1 (2.2)\\
$K^{*}(892)$ polarization                ~~~~&  6.1 (5.0)       ~~~~& 7.1 (6.5)      ~~~~& 6.6 (5.7)\\
\hline
\hline
Sum                                   ~~~~&  13.6 (14.0)      ~~~~& 14.5 (14.4)     ~~~~& 13.7 (14.0)\\
\hline
\end{tabular}
\end{center}
\end{table*}

\begin{table*}
\caption{\label{table::sys error} Systematic uncertainties on the branching fraction for the
$\chi_{cJ}\rightarrow\phi K^+K^-\pi^0$ final states (in \%).}
\begin{center}
\begin{tabular}{lccc}
\hline
Sys. err source          &  $\chi_{c0}$  & $\chi_{c1}$ & $\chi_{c2}$ \\
\hline
\hline
\multicolumn{4}{c}{Common contribution}\\
\hline
$\psp$ total number            &\multicolumn{3}{c}{ 0.8 }\\
Tracking                  & \multicolumn{3}{c}{$6.3$} \\
PID                           & \multicolumn{3}{c}{$8.9$} \\
Photon selection          &\multicolumn{3}{c}{$1\times3$} \\
$\pi^0$ reconstruction   & \multicolumn{3}{c}{1} \\
Kinematic Fit                   &\multicolumn{3}{c}{ 1} \\
\hline
$\phi$ mass window       ~~~~&  0.5        ~~~~& 1.2       ~~~~& 2.2 \\
$K^*$ mass window        ~~~~&  0.5        ~~~~& 2.7       ~~~~& 2.6 \\
$\chi_{cJ}$ lineshape    ~~~~&  8.5        ~~~~& 9.3       ~~~~& 6.2 \\
Fit range                             ~~~~&  2.4        ~~~~& 1.1       ~~~~& 1.0 \\
Non-peaking BG             ~~~~&  3.3        ~~~~& 1.6       ~~~~& 0.7 \\
Peaking BG without $\phi$ ~~~~&  0.5        ~~~~& 0.6       ~~~~& 0.4 \\
Peaking BG with $\phi$    ~~~~&  1.9        ~~~~& 4.3       ~~~~& 2.0 \\
BRs from PDG           ~~~~&  2.7        ~~~~& 3.3       ~~~~& 3.4 \\
MC model                 ~~~~&  2.0        ~~~~& 1.0       ~~~~& 0.9 \\
$K^{*}(892)$ polarization ~~~~& 6.6        ~~~~& 6.2       ~~~~& 7.4 \\
\hline
\hline
Sum                      ~~~~&  16.8       ~~~~& 17.3      ~~~~& 15.9 \\
\hline
\end{tabular}
\end{center}
\end{table*}

\section{\label{sec:mKKpi} the $\kkbarp$ invariant mass and $h_1(1380)$ state}

\begin{figure*}[htbp]
    \begin{center}
    \begin{overpic}[width=0.45\textwidth,angle=0]{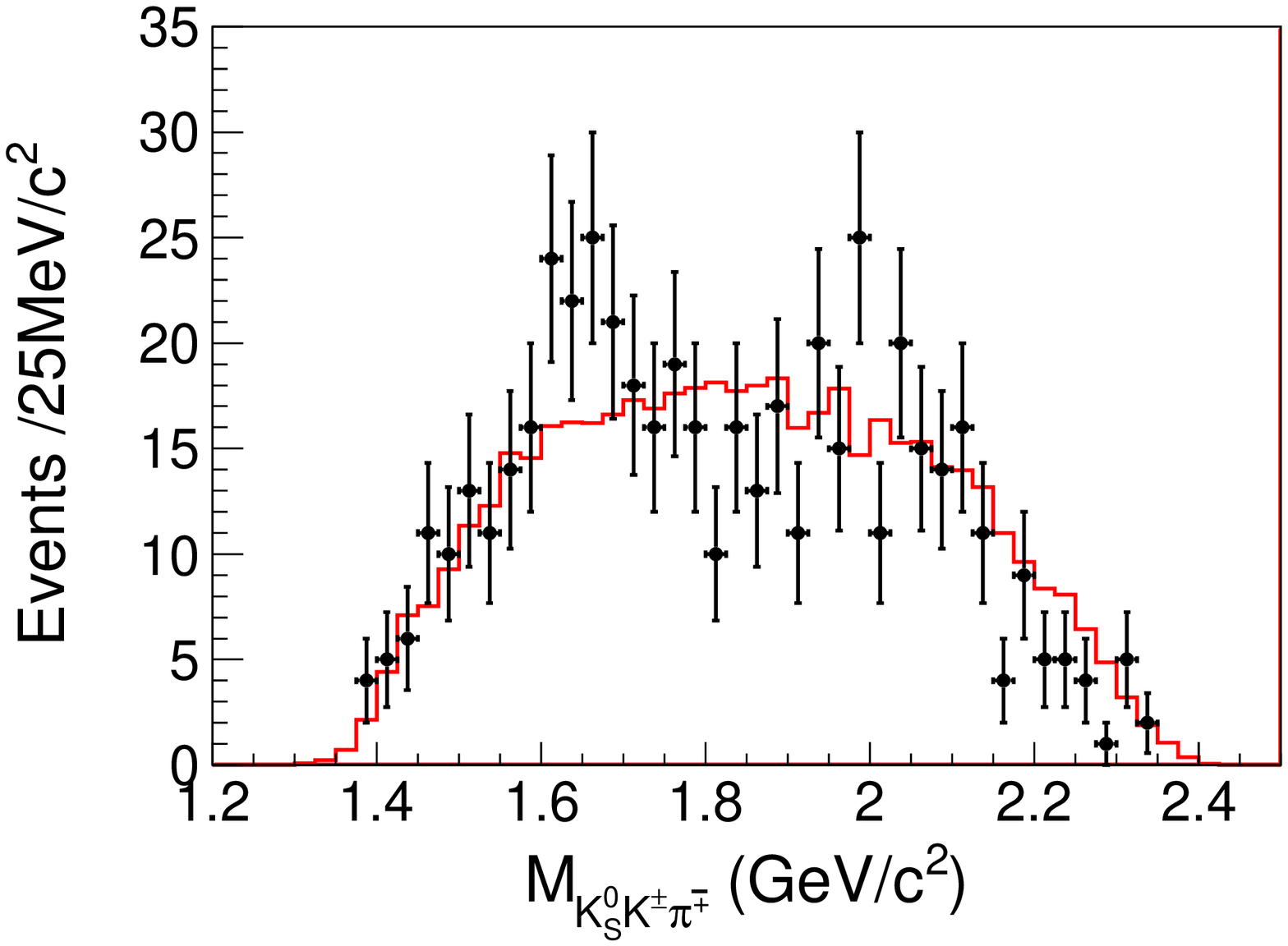}
     \put(80,50){(a)}
      \end{overpic}
    \begin{overpic}[width=0.45\textwidth,angle=0]{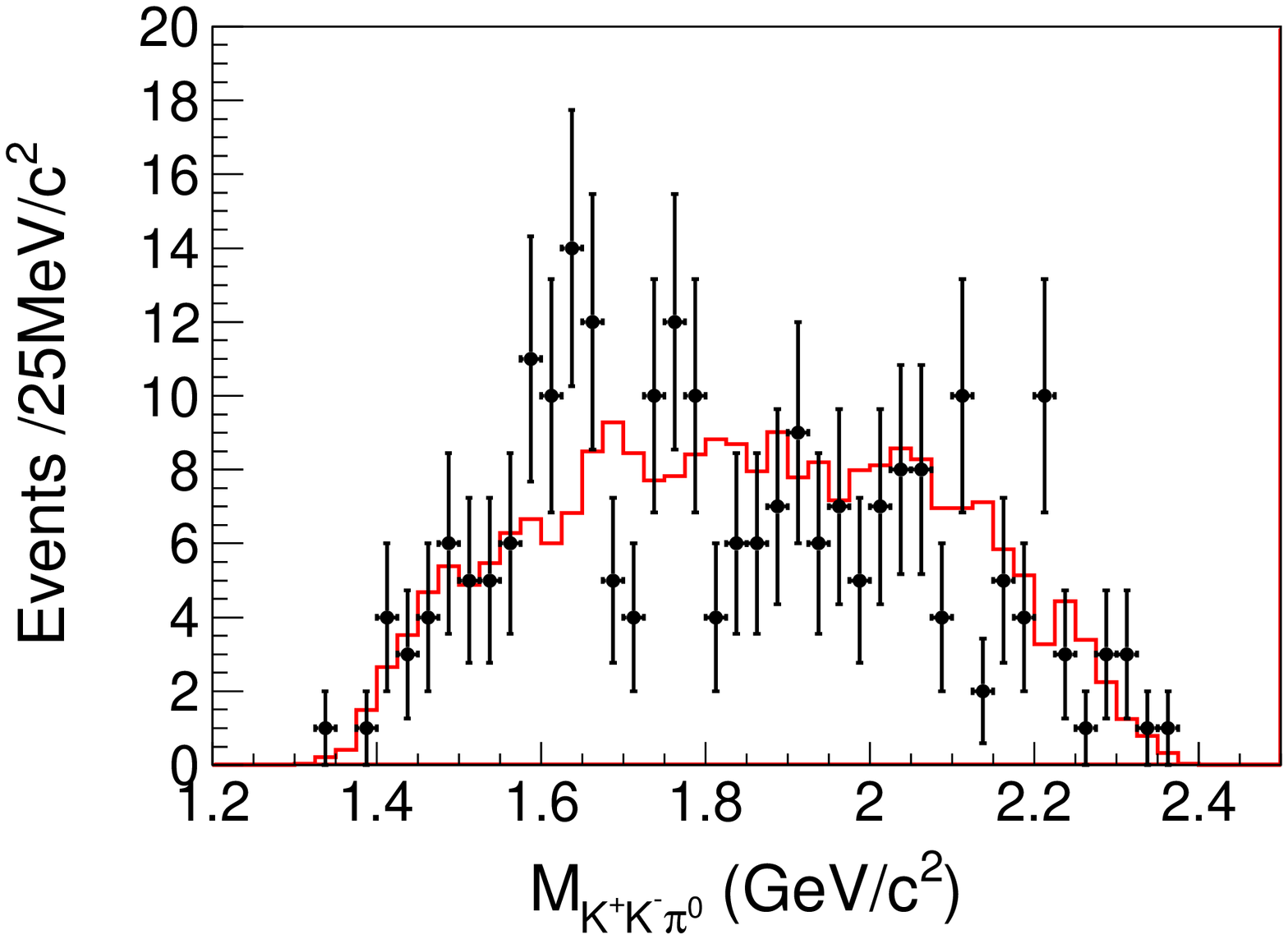}
    \put(80,50){(d)}
     \end{overpic}
     \begin{overpic}[width=0.45\textwidth,angle=0]{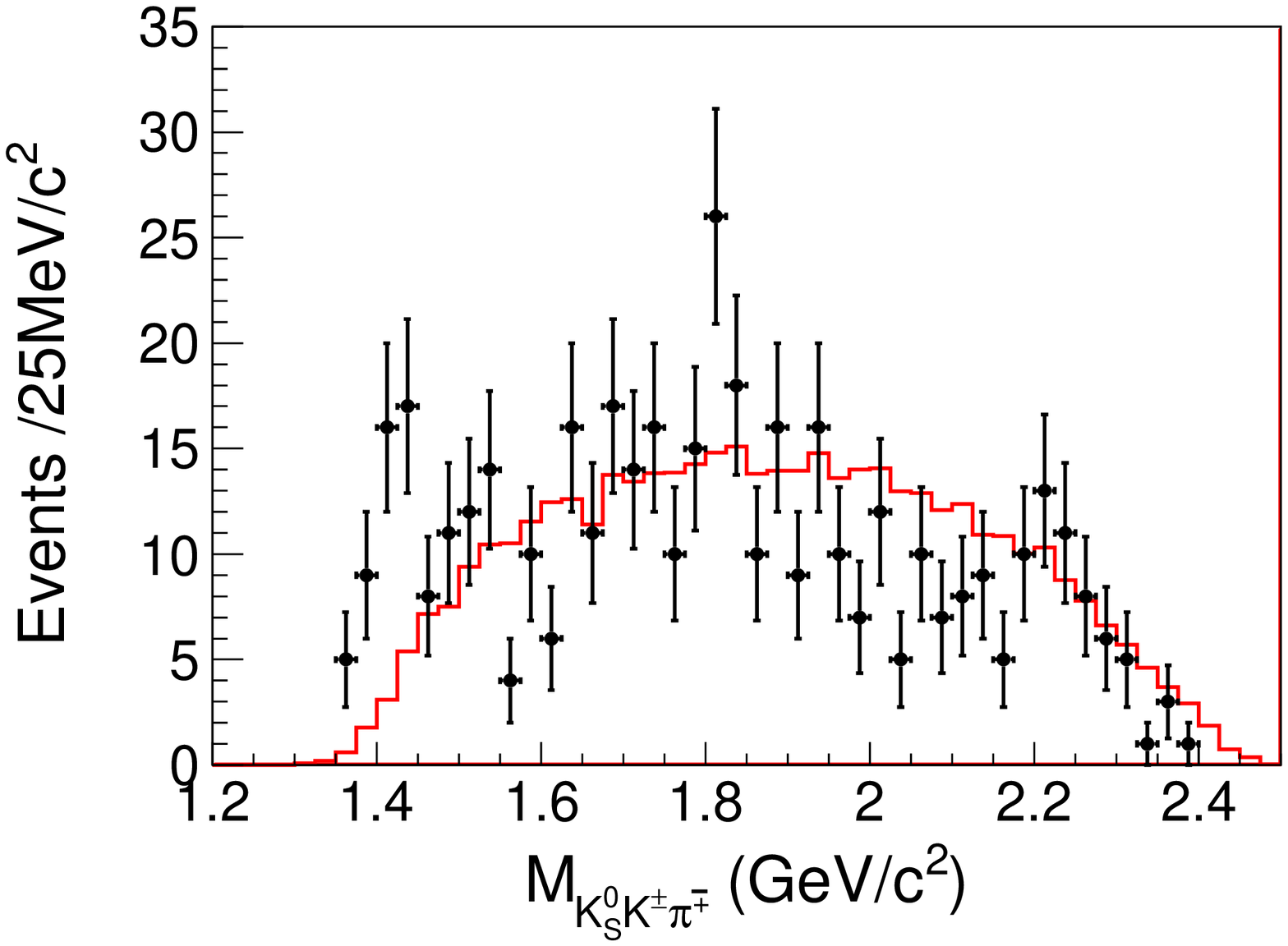}
     \put(80,50){(b)}
      \end{overpic}
    \begin{overpic}[width=0.45\textwidth,angle=0]{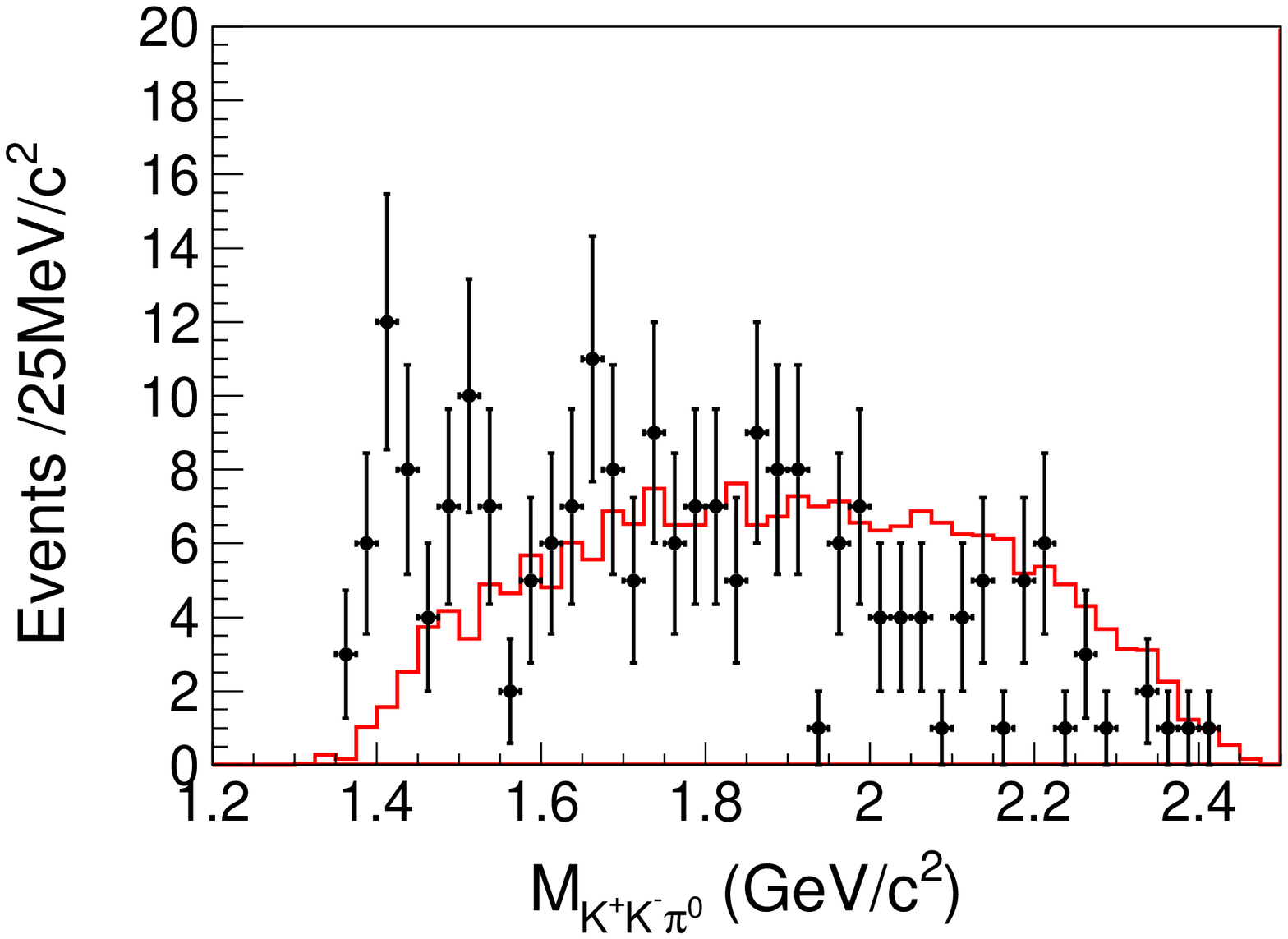}
    \put(80,50){(e)}
     \end{overpic}
     \begin{overpic}[width=0.45\textwidth,angle=0]{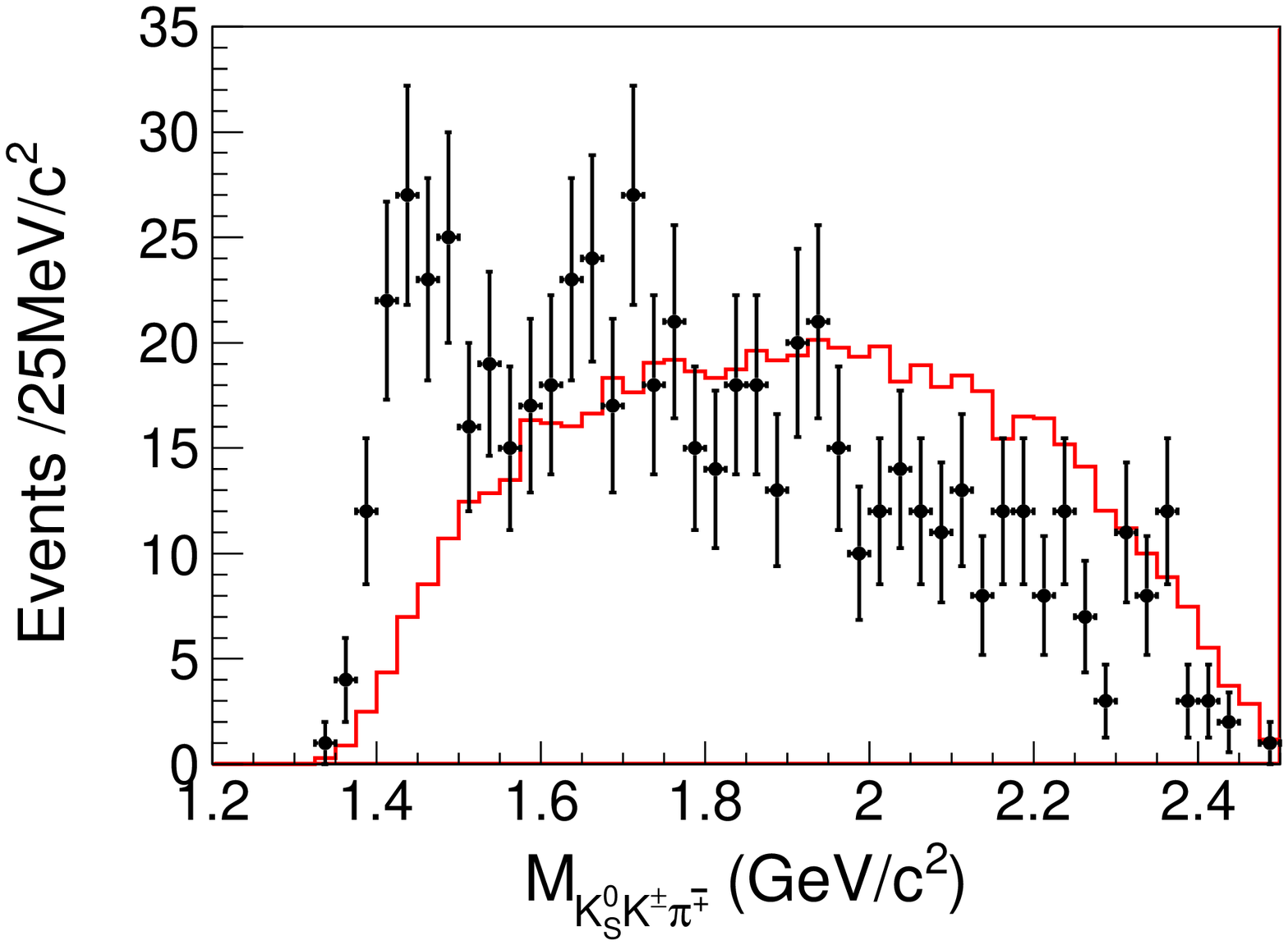}
     \put(80,50){(c)}
      \end{overpic}
    \begin{overpic}[width=0.45\textwidth,angle=0]{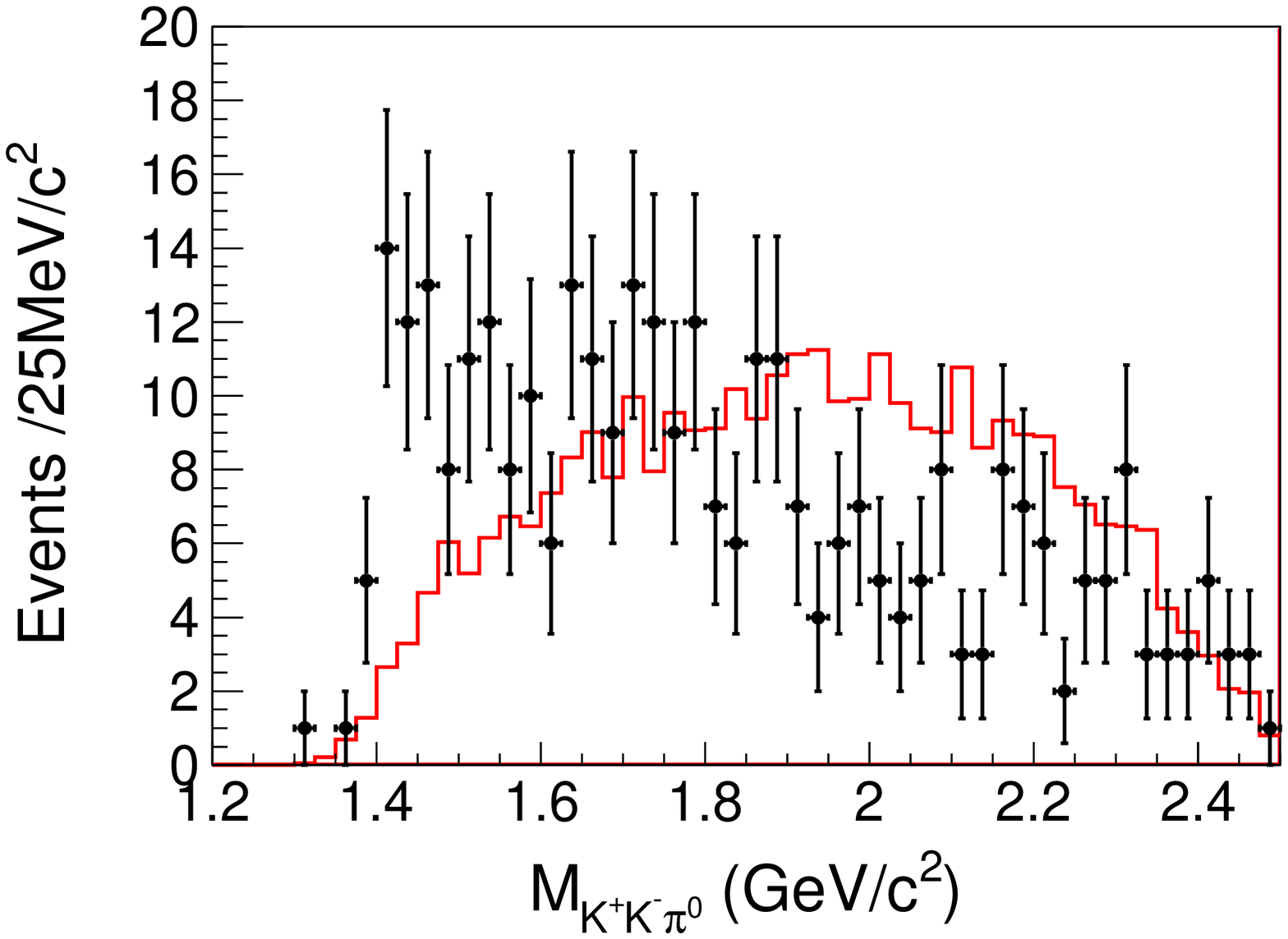}
    \put(80,50){(f)}
     \end{overpic}
    \end{center}
    \caption{
      The $\kkbarp$ invariant mass distributions, (a) $\kskp$ in the $\chi_{c0}$ region; (b) $\kskp$ in the $\chi_{c1}$ region; (c) $\kskp$ in the $\chi_{c2}$ region;  (d) $\kkp$ in the $\chi_{c0}$ region; (e) $\kkp$ in the $\chi_{c1}$ region; (f) $\kkp$ in the $\chi_{c2}$ region. The black dots represent the data, and the histogram shows the phase space MC sample.}
    \label{fig:kkp6}
\end{figure*}

The $\kkbarp$ invariant mass distributions are studied in order to identify any intermediate states.
Figure~\ref{fig:kkp6} shows the distributions of the $\kkbarp$ mass for the candidate events within the
$\chi_{0,1,2}$ signal regions of the data as well as the corresponding phase space MC samples
$\ccj\to\phi\kstark$. The $\chi_{c0,1,2}$ signal regions are defined as [3.365, 3.455] $\gevcc$,
[3.490, 3.530] $\gevcc$ and [3.540, 3.575] $\gevcc$, respectively.
A threshold enhancement, which can not be described with the phase space, is observed
in both $\cco$ and $\cct$ signal regions (Fig.~\ref{fig:kkp6} (b), (c), (e), (f)), but is absent
in the $\ccz$ signal region (Fig.~\ref{fig:kkp6} (a), (d)).

Assume that the threshold enhancement comes from an intermediate state, called X.
Since the X state is produced in the $\chi_{c1,2}$ decay associated with a $\phi$ meson, its $C$
parity is must be negative due to the conservation of $C$ parity.
The X state is observed in the $\kstark$ final state, and in the $\cct$ decay process
associated with a $\phi$ meson, which rules out an assignment of $J^P=0^+$.
If we assume that the X state is a conventional $q\bar{q}$ state, we can also rule out a $J^P$ of $0^-$,
$2^+$ {\it etc}. Taking into account its mass region, its decay through $\kstark$, and the accessible
$J^{PC}$ and comparing with the PDG~\cite{pdg}, the
most likely assignment for X is the $h_1(1380)$ ($J^{PC} = 1^{+-}$).
In the following, we assume that the $\kstark$ threshold enhancement is the $h_1(1380)$ intermediate state.

Besides the $\kstark$ threshold enhancement observed in the $\chi_{c1,2}$ signal regions, a clear
structure around 1.7 $\gevcc$ is observed in the $\ccz$ decay, also evident in the $\chi_{c1,2}$ decay
(Fig.~\ref{fig:kkp6}).
Considering  the mass region, and its decay
through $\kstark$, we conclude that this structure is the $\phi(1680)$.
The C parity should be minus, and the $\phi(1680)$ meets this requirement.
There is also a possible $\phi(1850)$ contribution visible in Fig.~\ref{fig:kkp6}.

\begin{figure}[h]
    \begin{center}
    \includegraphics[width=7.0cm,height=5.0cm,angle=0]{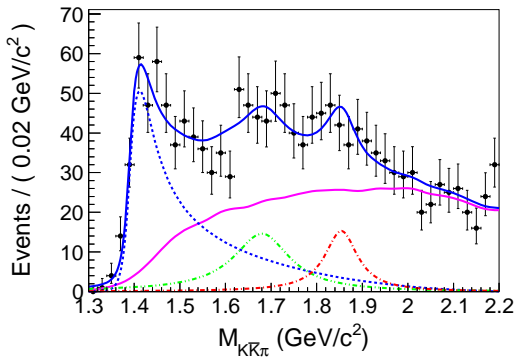}
    \end{center}
    \caption{The sum of
$K_S^0K^\pm\pi^\mp$ and $K^{\pm}K^{\mp}\pi^{0}$ mass spectra in the $\chi_{c1}$ and $\chi_{c2}$
mass regions. The markers with error bars represent the data; the dash curve the $h_{1}(1380)$ signal;
the dash-dot-dot curve the $\phi(1680)$ signal; and the dash-dot curve the $\phi(1850)$ signal.
    }
     \label{fig:h1380sumfit}
\end{figure}

To extract the resonance parameters of the $h_1(1380)$, a simultaneous fit is performed to the $\kskp$
and $\kkp$ invariant mass distributions for the $\chi_{c1,2}$ candidate events (Fig.~\ref{fig:kkp6}
(b), (c), (e), (f)).
Three resonance states, $h_1(1380)$, $\phi(1680)$ and $\phi(1850)$ as well as a phase space
contribution from $\chi_{c1,2}\to\phi\kstark$ are included in the fit without interference.
The phase space distribution is described using the shape of the MC
samples, smoothed with the KEYS method \cite{keys}.
The shapes of $\phi(1680)$ and $\phi(1850)$ are described by
relativistic Breit-Wigner functions with constant widths. All the
parameters of the resonances are fixed to PDG values~\cite{pdg}.
Based on the results of the $\chi_{cJ}$ fits, the background from
non-$\chi_{cJ}$ contributes is about 2.7{\%}.  MC studies show that the shape of the
$\kkbarp$ invariant mass of the non-$\chi_{cJ}$ contribution is similar to that of the phase space. Therefore, this background
is included in the phase space contribution in the fit to the $\kskp$ and $\kkp$ invariant mass distributions.
Since the $h_1(1380)$ resonance is close to the $\kstark$ mass threshold, it is parameterized with
a relativistic Breit-Wigner function with a mass dependent width~\cite{relBW}.
A detailed description of the $h_1(1380)$ lineshape
used in the following fits is given in Appendix~\ref{sec:appA}.
The simultaneous fit to the $\kskp$ and $\kkp$ invariant mass distributions is performed for the
candidate events in the $\cco$ and $\cct$ mass regions. The fit yields a mass of ($1412.4\pm4.4$)
$\mevcc$ and a width of ($84.5\pm12.4$) $\mev$ for the $h_1(1380)$ resonance, where
the errors are statistical only.
In the fit, we take the change of the detection efficiency as a
function of the invariant mass into account.
The efficiency functions are $0.12 - 0.02\times M_{\kskp}, 0.14 - 0.03\times M_{\kskp}, 0.18 - 0.02\times M_{\kkp} $ and $0.18- 0.03\times M_{\kkp}$
for the $\cco$ and $\cct$ regions, respectively.
The product of the relativistic Breit-Wigner functions and the
efficiency functions  are used to describe the signal.
Figure~\ref{fig:h1380sumfit} shows the sum of the invariant mass distribution of the $\kskp$ and
$\kkp$ decay modes for the candidate events in the $\cco$ and $\cct$ mass regions as well
as the sum of the corresponding fit curves.
The goodness of fit is determined to be $\chi^{2}/n.d.f = 1.09$ by projecting all
candidate events in 45 bins.
The statistical significance of the $h_1(1380)$ signal is measured to be greater than
10 $\sigma$ by comparing the likelihood values of the fit with and without the $h_1(1380)$.
The statistical significances of the $\phi(1680)$ and $\phi(1850)$ signals evaluated by the same
method are found to be 4.3$\sigma$ and 3.2$\sigma$, respectively.
We consider the interferences between $h_1(1380)$ and these resonances in the systematic uncertainties.

The different sources of systematic uncertainty for the $h_1(1380)$ resonance
parameters are considered as follows:

{\it (a) Parameterization of the energy dependent width. }
The energy dependent width of the $h_1(1380)$ is described with the
truncation functions combined of exponential and polynomial functions.
The curves of $S(m)$ are re-fitted with the same truncation functions, but
with different truncation point. 
A fit was also attempted to the $S(m)$ curves using a second-order polynomial.
The new fitted functions of $S(m)$ are used in the simultaneous fit on the $\kkbarp$
invariant mass. The largest differences in the mass and width of the $h_1(1380)$,
3.9 $\mevcc$ and 3.8 $\mev$, are taken as one of systematic uncertainties for the $h_1(1380)$
resonance parameters.

{\it (b) Background shape. }
In the fit, the background is described by a smoothed phase space MC sample.
An alternative fit is performed using a second-order Chebychev polynomial
function to describe the non-resonant. The differences in mass, 4.7 $\mevcc$, and width, 5.5 $\mev$,
are taken as the systematic uncertainties due to the shape of the background.

{\it (c) Fit range. }
An alternative fit with a different fit range, [1.2, 2.0] $\gevcc$, is performed.
The differences in the mass and width of the $h_1(1380)$ are found to be 0.5 $\mevcc$ and  3.5 $\mev$, respectively.

{\it (d) Efficiency parameterization. }
A fit with a flat efficiency curve is performed, and the differences of  0.3 $\mevcc$
in the mass and 1.0 $\mev$ in the width are taken as the
systematic uncertainties related to the efficiency correction.

{\it (e) $\phi(1680)$ resonance. }
In the nominal fit,
the mass and width of the $\phi(1680)$ resonance are fixed to their PDG values.
Different fit scenarios, (1) without the $\phi(1680)$, (2) leaving the mass
and width of the $\phi(1680)$ resonance free in the fit, and (3) changing the mass and width by one standard
deviation from the PDG values are performed, and the largest change in the mass and width, 1.4 $\mevcc$
and 17.0 $\mev$, are regarded as the systematic uncertainties related
to the $\phi(1680)$ resonance.

{\it (f) $\phi(1850)$ resonance. }
The systematic uncertainty due to the $\phi(1850)$ resonance is evaluated in a similar way as for the $\phi(1680)$.
The largest change in mass and width, 1.1 $\mevcc$ and 3.5 $\mev$, are taken as the systematic
uncertainties related to $\phi(1850)$ resonance.

{\it (g) The branching fraction constraint between isospin conjugate decay modes. }
In the nominal fit, no constraint is imposed on branching fractions
for the the isospin conjugate modes and different final states.
The systematic uncertainty related to the branching fraction constraint is estimated by requiring
that the branching fraction of $h_1(1380) \to \kskp$ is twice that of
$h_1(1380) \to \kkp$, which is expected by isospin symmetry. The changes in the mass and width
of the $h_1(1380)$, 0.3 $\mevcc$ and 3.4 $\mev$, are taken as the systematic uncertainties of this type.

{\it (h) Interference effects. }
The fit is repeated under the following interference scenarios,
(1) the interference between $h_1(1380)$ and phase space,
(2) the interference between $h_1(1380)$ and $\phi(1680)$,
(3) the interference between $h_1(1380)$ and $\phi(1850)$,
The largest differences to the nominal values of the mass and width,
4.7 $\mevcc$ and 35.3 $\mev$, are taken as the systematic
uncertainties related to interference effects.

Table ~\ref{table::error:h1380} shows the systematic
uncertainties for the mass and width of $h_1(1380)$.
Assuming that all sources of systematic uncertainty are independent, the total systematic uncertainty
is determined to be 8 $\mevcc$ for the mass and 40 $\mev$ for the width.

\begin{table*}
\caption{\label{table::error:h1380} Systematic uncertainties on mass and width for the
$h_1(1380)$.}
\begin{center}
\begin{tabular}{lcc}
\hline
Sys. err source          & mass ($\mev$)  ~~~~& width($\mev$) \\
\hline
\hline
Parameterization of energy dependence width            ~~~~& 3.9       ~~~~& 3.8 \\
Shape of the background           ~~~~& 4.7       ~~~~& 5.5 \\
Fit range         ~~~~& 0.5       ~~~~& 3.5 \\
Efficiency curve                                   ~~~~& 0.3       ~~~~& 1.0 \\
$\phi(1680)$ resonance                  ~~~~& 1.4       ~~~~& 17.0 \\
$\phi(1850)$ resonance        ~~~~& 1.1       ~~~~& 3.5 \\
Isospin constraint       ~~~~& 0.3       ~~~~& 3.4 \\
Interference effects             ~~~~& 4.7       ~~~~& 35.3 \\

\hline
\hline
Sum                           ~~~~& 7.9      ~~~~& 40.2 \\
\hline
\end{tabular}
\end{center}
\end{table*}

\section{\label{sec:sum}Summary and conclusions}
Using 106 M $\psp$ events collected with the BESIII detector, we present a study
of the decays $\ccj\to\phi\kskp$ and $\ccj\to\phi\kkp$, via the E1 radiative
transition $\psp\to\gamma\ccj$.
The decays are dominated by the three-body reaction $\ccj\to\phi\kstark$. The
branching fractions for this reaction via neutral and charged $\kstar$ are measured for
the first time and are summarized in Tables \ref{table:brfractionkskpisum}.
The measured branching fractions for $\ccj\to\phi\kstarkc$ in the two different final states are found to be consistent with each other.
The ratio of the branching fraction for $\ccj\to\phi\kstarkc$ to that
of $\ccj\to\phi\kstarkn$ is consistent within the expectations of
isospin symmetry.

By examining the invariant mass spectrum of $\kkbarp$, a significant
excess of events above the phase space expectation is observed
near the $\kstark$ mass threshold in the decays of $\chi_{c1,2}$, with a significance greater than 10$\sigma$.
The observed structure has negative $C$ parity, and is expected to
be the $h_1(1380)$ state, considering its mass, width and decay through $\kstark$.
A simultaneous fit is performed to the invariant mass distributions of $\kkbarp$ for the
candidate events in the $\chi_{c1,2}$ signal regions. The mass and width of the $h_1(1380)$ are determined to be
$1412\pm4(stat.)\pm8(sys.)~\mevcc$ and $84\pm12(stat.)\pm40(sys.)~\mev$, respectively.
This is the first direct observation of the $h_1(1380)$ in its decay to $\kstark$.
Evidence is also found for the decays $\ccj\to\phi\phi(1680)$ and $\ccj\to\phi\phi(1850)$, but
with significances less than 5$\sigma$.
More data and advanced analysis techniques, $e.g.$ PWA, may
shed light on the properties of the structures observed in the
$\kkbarp$ invariant mass spectrum.

\begin{table*}[htbp]
\caption{\label{table:brfractionkskpisum} Branching fractions
measured in $\phi \kkbarp$ final states}
\begin{center}
\begin{tabular}{cccc}
\toprule
\multicolumn{2}{c}{Decay Modes}                & $\phi K_s K^\pm \pi^\mp$ ($\times 10^{-3}$)             &$\phi K^+ K^- \pi^0$ ($\times 10^{-3}$)\\
\hline
\multirow{2}{*}{$\chi_{c0}$} & $\phi K^{*}(892)^{\pm}K^{\mp}$         ~~~~& $1.65\pm0.21(stat.)\pm0.22(sys.)$         ~~~~&$1.90\pm0.14(stat.)\pm0.32(sys.)$\\
                                                      & $\phi K^{*}(892)^{0}\bar{K^0}+c.c.$    ~~~~& $2.03\pm0.21(stat.)\pm0.28(sys.)$         ~~~~&---\\
\hline
\multirow{2}{*}{$\chi_{c1}$} & $\phi K^{*}(892)^{\pm}K^{\mp}$          ~~~~& $1.76\pm0.21(stat.)\pm0.26(sys.)$          ~~~~&$1.62\pm0.12(stat.)\pm0.28(sys.)$\\
                                                      & $\phi K^{*}(892)^{0}\bar{K^0}+c.c.$     ~~~~& $1.51\pm0.19(stat.)\pm0.22(sys.)$          ~~~~&---\\
\hline
\multirow{2}{*}{$\chi_{c2}$} & $\phi K^{*}(892)^{\pm}K^{\mp}$           ~~~~& $2.56\pm0.23(stat.)\pm0.35(sys.)$          ~~~~&$2.74\pm0.16(stat.)\pm0.44(sys.)$ \\
                                                      & $\phi K^{*}(892)^{0}\bar{K^0}+c.c.$      ~~~~& $2.27\pm0.22^(stat.)\pm0.32(sys.)$       ~~~~&---\\
\hline
\end{tabular}
\end{center}
\end{table*}

\acknowledgments
The BESIII collaboration thanks the staff of BEPCII and the IHEP computing center for their strong support. This work is supported in part by National Key Basic Research Program of China under Contract No. 2015CB856700; National Natural Science Foundation of China (NSFC) under Contracts Nos. 10935007, 11121092, 11125525, 11235011, 11322544, 11335008, 11375170, 11275189, 11079030, 11475164, 11005109, 11475169;  the Chinese Academy of Sciences (CAS) Large-Scale Scientific Facility Program; Joint Large-Scale Scientific Facility Funds of the NSFC and CAS under Contracts Nos. 11179007, U1232201, U1332201; CAS under Contracts Nos. KJCX2-YW-N29, KJCX2-YW-N45; 100 Talents Program of CAS; INPAC and Shanghai Key Laboratory for Particle Physics and Cosmology; German Research Foundation DFG under Contract No. Collaborative Research Center CRC-1044; Istituto Nazionale di Fisica Nucleare, Italy; Ministry of Development of Turkey under Contract No. DPT2006K-120470; Russian Foundation for Basic Research under Contract No. 14-07-91152; U. S. Department of Energy under Contracts Nos. DE-FG02-04ER41291, DE-FG02-05ER41374, DE-FG02-94ER40823, DESC0010118; U.S. National Science Foundation; University of Groningen (RuG) and the Helmholtzzentrum fuer Schwerionenforschung GmbH (GSI), Darmstadt; WCU Program of National Research Foundation of Korea under Contract No. R32-2008-000-10155-0.

\appendix
\section{Appendix: Lineshape of $h_1(1380)$}
\label{sec:appA}
The $h_1(1380)$ resonance is parameterized with
a relativistic Breit-Wigner function with a mass dependent width
\begin{equation}
\frac{m_0^{2}\Gamma^{2}(m;m_0)}{(m_0^2-m^2)^2+m_0^2\Gamma^2(m;m_0)},
\end{equation}
where $m_0$ is the nominal mass of $h_1(1380)$ state, and $\Gamma(m;m_0)$ is the corresponding
mass dependent width, which will be discussed in the following.

In typical two-body decays, the width $\Gamma(m, m_{0})$ varies with mass roughly as~\cite{twobodywidth}
\begin{equation}
\Gamma(m, m_0) \approx \Gamma_0(\frac{q}{q_0})^{2l+1},
\end{equation}
where $l$ is the orbital angular momentum, $q$ is the momentum available in a normal two-body
decay, and $\Gamma_0$ and $q_0$ are the corresponding widths and momenta in the nominal mass.
However, in a quasi two-body decay such as $h_1(1380)\to\kstark$, the $q$ of the decay isobar
($\kstar$) is no longer precisely defined. Due to the non-zero $\kstar$ width, the
$\kstark$ threshold is not well defined and the $q$ momentum available in a normal two-body amplitude
may become un-physical ($q^2\leq0$).
Following Ref.~\cite{Emeson}, if we assume the energy dependence of $\Gamma$ is proportional
to the integral of the decay matrix element over the available area of the Dalitz plot for
each $K\bar{K}\pi$ mass value, then
\begin{widetext}
\begin{equation}
\Gamma(m;m_0) =
\Gamma_0\frac{S(m)}{S(m_0)}=
\Gamma_0\frac{\int\limits_{Dalitz ~plot
(m)}D_{K^*}^2d\Phi}{\int\limits_{Dalitz ~plot (m_0)}D_{K^*}^2d\Phi},
\end{equation}
\end{widetext}
and for $S$-wave $\kstark$
\begin{equation}
D_{K^*} = |BW_{K^*_{12}}\cdot\vec{t}_3+g\cdot BW_{K^*_{13}}\cdot\vec{t}_2|,
\end{equation}
where $BW_{K^*_{ij}}$ are the usual $P$-wave relativistic Breit-Wigner functions~\cite{twobodywidth}
with the mass and width of $\kstar$ fixed at their PDG values. Here $\vec{t}_i$ are the Zemach vectors
describing the spin 1 content, and are the vector part of
\begin{equation}
p_j^\mu-p_k^\mu-[\frac{m_j^2-m_k^2}{m_{jk}^2}][p_j^\mu+p_k^\mu],
\end{equation}
and $\vec{p_i}$ (i = 1, 2, 3) are the vector parts of the 4-momenta of $\pi$, $K$ and $\bar{K}$
in the $\kkbarp$ rest system.
The interference between the $\kstark$ and $\bar{K}^*(892)K$ is fixed by $C$ parity:
constructive ($g = +1$) for $C = +1$, and destructive ($g = -1$) for $C = -1$. In our analysis, $C = -1$.

To simplify the fitting process, a numerical method is used to extract the integral
of the decay matrix element over the available phase space.
The integral can be re-written as :
\begin{widetext}
\begin{equation}
S(m)\equiv\int\limits_{Dalitz~plot (m)}D_{K^*}^2d\Phi \simeq
\frac{V}{N}\sum_{i=1}^N D_{K^*}^2(i)
=V\frac{1}{N}\sum_{i=1}^N D_{K^*}^2(i),
\end{equation}
\end{widetext}
where the term $\frac{1}{N}\sum_{i=1}^N D_{K^*}^2(i)$ can be extracted using the
MC samples of $\chi_{cJ}\to\phi X$ $(X\to\kkbarp)$ with a large statistics N,
and V is the phase space volume for the decay $X\to\kkbarp$.

A set of MC samples generated with a different mass $m$ and zero width for the intermediate
state X is used to extract $\frac{1}{N}\sum_{i=1}^N D_{K^*}^2(i)$.
The phase space volume V of a three-body decay with system mass $m$ can be derived as~\cite{FJames}:
\begin{widetext}
\begin{equation}
R_3(m;m_1,m_2,m_3) = \int_{(m_1+m_2)^2}^{(m-m_3)^2} R_2(m;m_{12},m_3)R_2(m12;m_1,m_2)dm_{12}^2
\end{equation}
\end{widetext}
where $m_{12}$ is the invariant mass of system of particle 1 and particle 2, and $R_2$ is the
two-body phase space volume, which can directly taken from PDG~\cite{pdg} :
\begin{widetext}
\begin{equation}
R_2(m_{12};m_1,m_2) =
\sqrt{[1-(\frac{m_1+m_2}{m_{12}})^2][1-(\frac{m_1-m_2}{m_{12}})^2]}
\end{equation}
\end{widetext}
The above integral process gives the same result as the integral over the decay area in the
Dalitz decay study~\cite{dalitzdecay}.

\begin{figure}[h]
    \begin{center}
    \begin{overpic}[width=7.0cm,height=5.0cm,angle=0]{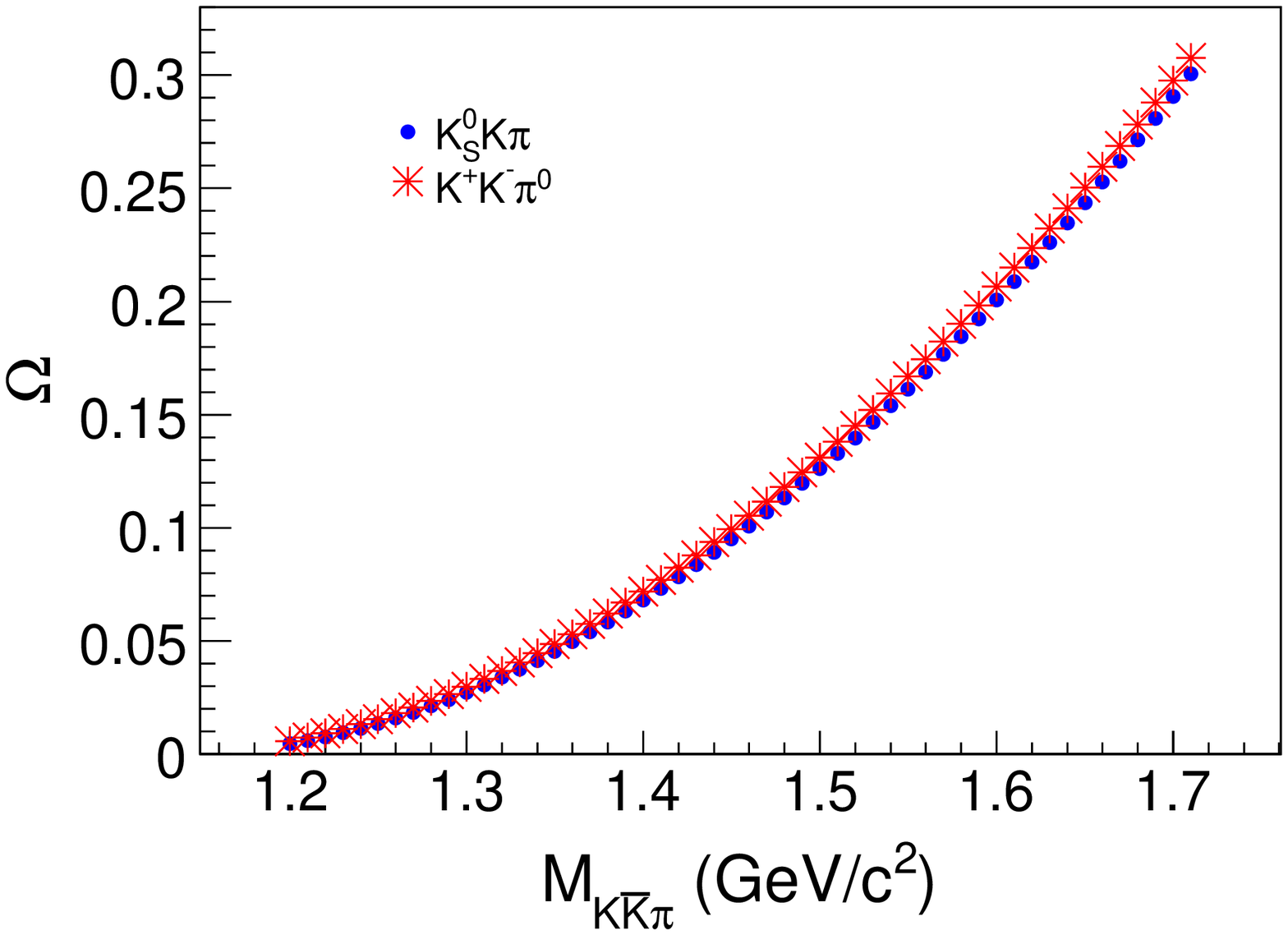}
    \put(20,50){(a)}
    \end{overpic}

    \begin{overpic}[width=7.0cm,height=5.0cm,angle=0]{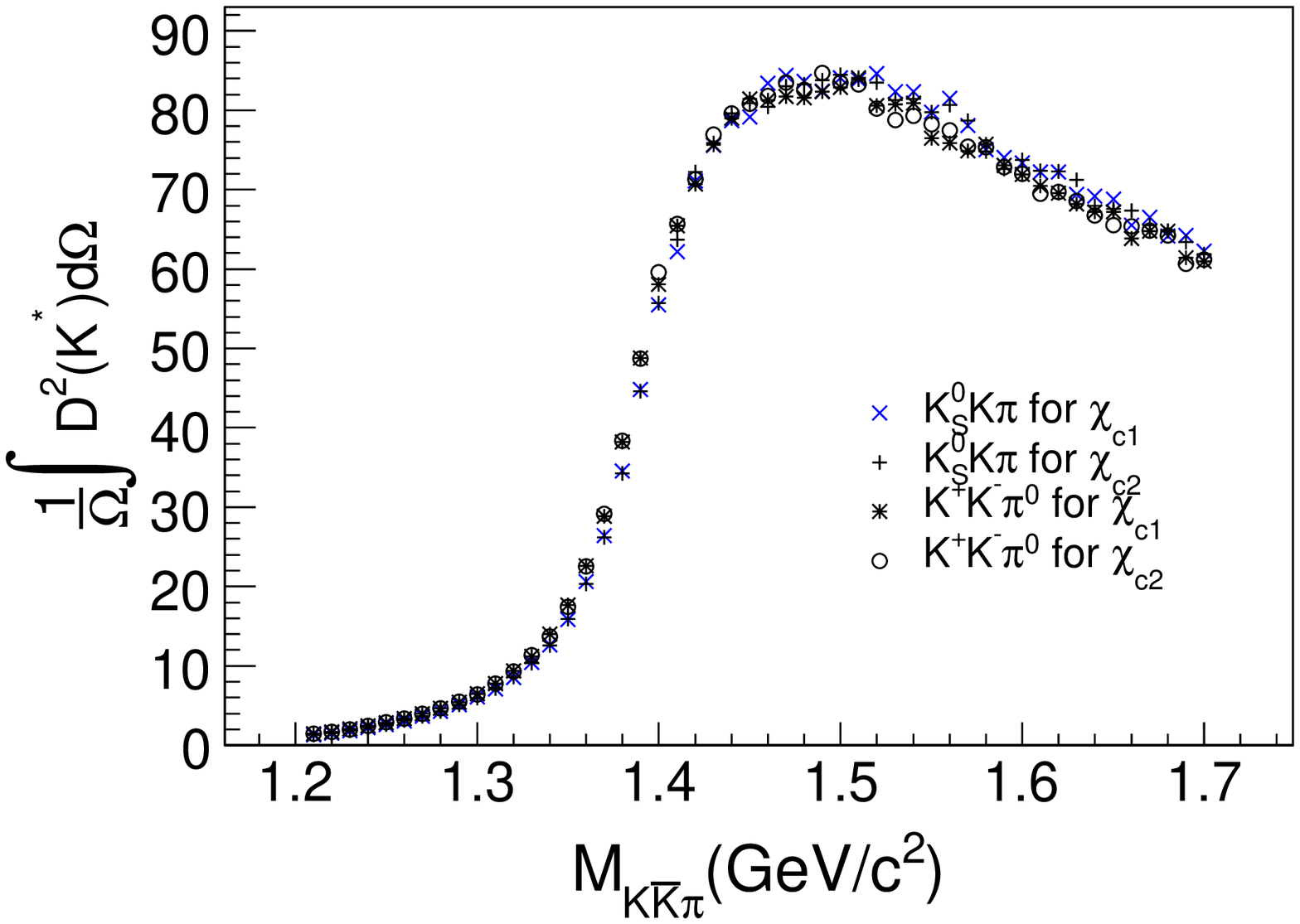}
    \put(20,50){(b)}
    \end{overpic}

    \begin{overpic}[width=7.0cm,height=5.0cm,angle=0]{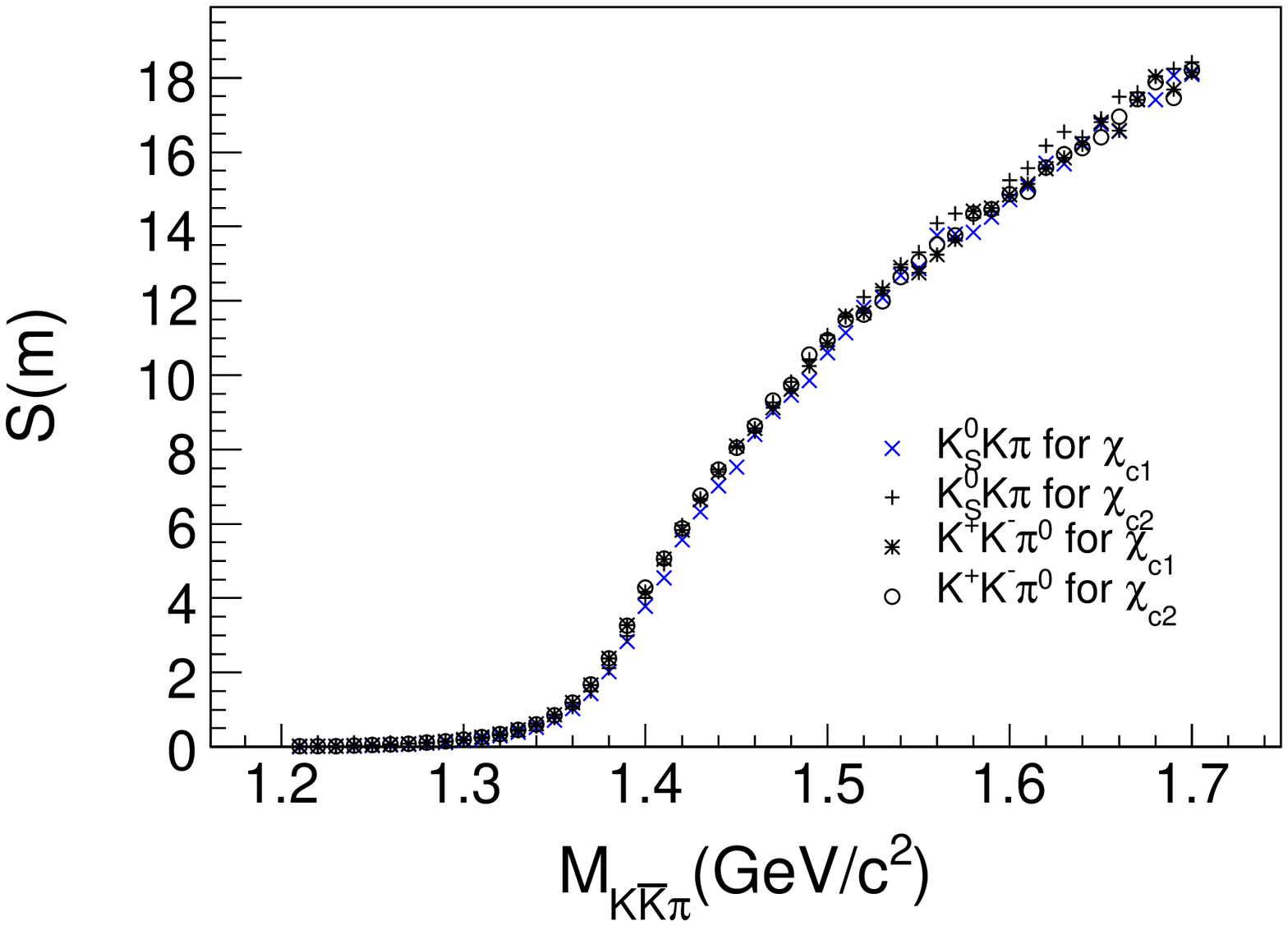}
    \put(20,40){(c)}
    \end{overpic}
    \end{center}
    \caption{
         Numerical calculation of S(m),
        (a) phase space volume V, (b) $\frac{1}{N}\sum_{i=1}^N D_{K^*}^2(i)$,
        (c) S(m).
     }
    \label{fig:cutss}
\end{figure}

The value of $\frac{1}{N}\sum_{i=1}^N D_{K^*}^2(i)$, the phase space volume V, and the
corresponding product, $S(m)$, as function of invariant mass of $\kkbarp$ system, $m$,
are shown in Fig.~\ref{fig:cutss}.
Due to the small difference in the mass and width of neutral and charged $\kstar$ and  $\chi_{c1,2}$
phase space volume, the calculations are performed for $\chi_{c1,2}$ and the decay mode of $\kskp$ and $\kkp$, individually.
$S(m)$ represents the mass dependence of the width $\Gamma(m, m^{0})$ for the $h_1(1380)$, which is used to describe the
$h_1(1380)$ line-shape.

\end{document}